\documentclass[aps,twocolumn,showpacs]{revtex4-1}
\usepackage{graphicx}
\usepackage{amsmath,amsfonts}
\usepackage[T1]{fontenc}
\usepackage[usenames]{color}

\begin{document}

\title{Thermodynamics of a spin-1 Bose gas with fixed magnetization}

\author{Guillaume Lang}
\affiliation{Instytut Fizyki PAN, Aleja Lotnik\'ow 32/46, 02-668 Warsaw, Poland}

\author{Emilia Witkowska}
\affiliation{Instytut Fizyki PAN, Aleja Lotnik\'ow 32/46, 02-668 Warsaw, Poland}

\begin{abstract}
We investigate the thermodynamics of a spin-1 Bose gas with fixed magnetization including the quadratic Zeeman energy shift. 
Our calculations are based on the grand canonical description for the ideal gas and the classical fields approximation for atoms with ferromagnetic and antiferromagnetic interactions. We confirm the occurence of a double phase transition in the system that takes place due to two global constraints. 
We show analytically for the ideal gas how critical temperatures and condensed fractions are changed by a non-zero magnetic field. 
The interaction strongly affects the condensate scenario below the second critical temperature. 
The effect imposed by interaction energies becomes diminished in high magnetic fields where condensation, of both ferromagnetic and antiferromagnetic atoms, agree with the ideal gas results.
\end{abstract}
\pacs{
03.75.Mn, 
05.30.-d, 
05.70.Fh, 
03.50.-z, 
}

\maketitle

\section{Introduction}

A spinor Bose-Einstein condensate (BEC) is a multi-component condensate with an additional spin degree of freedom, which has provided exciting opportunities to study experimentally quantum magnetism, superfluidity, strong correlations, coherent spin-mixing dynamics, spin-nematic squeezing, entanglement etc, most of them in non-equilibrium situations (see ~\cite{{quantum_magnetism},{superfluidity},{strong_correlations},{spin_mixing},{spin_squeezing},{massive_entanglement}}). Despite successful experimental developments on spinor Bose-Einstein condensates, our knowledge remains limited regarding equilibrium properties and in particular the thermodynamics of such a gas. The main reason is a long time needed to reach an equilibrium state, typically several seconds or tens of seconds, that may exceed the lifetime of the condensate \cite{StamperKurn}. Nevertheless, recent experimental developments allowed for investigation of the ground state of an antiferromagnetic spinor condensate opening the paths to study in details its properties at thermal equilibrium~\cite{GS_experiment}.

The condensation of atoms with total spin $F=1$ trapped in the three hyperfine states $m_F=1,0,-1$ in the absence of magnetic field was investigated theoretically for the first time by Isoshima et al.~\cite{Isochima}. The double condensation phenomenon was predicted in the presence of two global conserved quantities: the total number of atoms $N$ and the magnetization $M$. A condensate starts to appear in the highest $m_F=1$ component for temperatures below the first critical temperature and simultaneously in the rest two components for temperatures below the second critical temperature. Analytical expressions for the two critical temperatures and condensate fractions were given for the ideal gas and zero magnetic field~\cite{Isochima, Kao}. The condensation of interacting spin-1 Bose gas was considered numerically within the Bogoliubov-Popov approximation~\cite{Isochima} and the Hartree-Fock-Popov approximation~\cite{Zhang}. In the latter, authors confirmed the double phase transition for antiferromagnetic interactions, but found a more complicated phase diagram for ferromagnetic interactions with a possible triple condensation scenario. The only one experimental work of Pasquiou~\cite{Laburthe} touches the problem of the thermodynamics in chromium atoms with total spin $F=3$ but for free magnetization. Indeed, for low magnetic fields when the magnetization is approximately conserved the experimental results confirm the occurrence of a double condensation. 

In this paper, we reconsider the topic of condensation in the system of spin-1 bosons with fixed magnetization. The ultra-cold gases are almost perfectly isolated in the experiment and conservation of magnetization plays a major role. The magnetic dipole-dipole interactions, that may change the magnetization, are realtively weak and can be neglected for $F=1$ sodium or rubidium spinor Bose-Einstein condensates. 

We examine the thermodynamics of the ideal gas in the presence of quadratic Zeeman effect within the grand canonical ensemble. A non-zero magnetic field introduces a new phase in the phase diagram of critical temperatures that we characterize by the threshold magnetization. The condensation scenario predicted by Isoshima is present for magnetizations larger than the thereshold magnetization. When the magnetization of the system is smaller than the threshold magnetization, atoms start condensing first not in the highest $m_F=1$ component, as it was the case for zero magnetic field, but in the $m_F=0$ component. That trivial effect is present due to the shift of the lowest energy level of the $m_F=0$ component below the lowest energy level of the $m_F=1$ component. We give an explicit expression for the threshold magnetization.

We study the interacting gas within the classical fields approximation~\cite{CFM} combined with the Metropolis algorithm~\cite{metropolis}. The method was successfully used to investigate thermal effects in the single-component Bose-Einstein condensates including thermodynamics~\cite{Optc}, vortex-dynamics~\cite{vortex}, critical temperature shift~\cite{Tcshift}, spin-squeezing~\cite{sqthermal}, solitons or Kibble-Zurek mechanism~\cite{solitons}, and many others, some of them reviewed in~\cite{ref_TH}. This numerical method includes all non-linear terms present in the Hamiltonian at the expense of introducing a free parameter that has to be well chosen. In this paper we explain how to adapt the Metropolis algorithm for a spin-1 gas with fixed magnetization. To demonstrate the validity of the proposed algorithm we compared results of simulations with exact results for the ideal gas and with the approximated Bogoliubov theory for antiferomagnetic interactions. We confirmed double condensation for both ferromagnetic and antiferromagnetic interactions. The condensation strongly differs from the result for ideal gas below the second critical temperature. In the high magnetic field limit, when the quadratic Zeeman energy dominates over the interaction energy, details of condensation do not depend on the interaction sign and are well described by the ideal gas results.

\section{The model}

We consider a dilute and homogeneous spin-1 Bose gas in a magnetic field.
We start with the Hamiltonian $H = H_0 + H_{\rm A}$, where the symmetric (spin-independent) part is
\begin{equation} \label{En}
H_0 = \sum_{j=-,0,+} \int d^3 r \, \psi_j^\dagger \left(-\frac{\hbar^2}{2m}\nabla^{2} + \frac{c_0}{2} n \right) \psi_j.
\end{equation}
Here the subscripts $j=-,0,+$ denote sublevels with magnetic quantum numbers along the magnetic field axis $m_F=-1,0,+1$,
$m$ is the atomic mass, $n=\sum n_j = \sum \psi_j^\dagger \psi_j$ is the total atom density.
The spin-dependent part can be written as
\begin{equation} \label{EA}
H_{\rm A} = \int d^3r \, \left[ \sum_j E_j n_j + \frac{c_2}{2} :{\bf F}^2:\right]\,,
\end{equation}
where $E_j$ are Zeeman energy levels, ${\bf F}=(\psi^{\dagger}f_x\psi,\psi^{\dagger}f_y\psi,\psi^{\dagger}f_z\psi)^T$ is the spin density,
$f_{x,y,z}$ are spin-1 matrices, $\psi =(\psi_+,\psi_0,\psi_-)^T$, and $:\,:$ denotes the normal order.
The spin-independent and spin-dependent interaction coefficients are given by
$c_0=4 \pi \hbar^2 (a_0+2 a_2)/3m$ and $c_2= 4 \pi \hbar^2 (a_2 - a_0)/3m$ respectively,
where $a_S$ is the s-wave scattering length for colliding atoms with total spin $S$.
The total number of atoms
\begin{equation}\label{N}
N = \int n d^3r
\end{equation}
and the magnetization
\begin{equation}\label{M}
M = \int \left(n_+ - n_-\right) d^3r
\end{equation}
are conserved quantities.

The linear part of the Zeeman shifts $E_j$ induces a homogeneous rotation of the spin vector around the direction of the magnetic field.
Since the Hamiltonian is invariant with respect to such spin rotations, we consider only the effect of the quadratic Zeeman shift.

For a sufficiently weak magnetic field we can approximate Zeeman energy levels by a positive energy shift of the $m_F=\pm 1$ sublevels
$\delta=(E_+ + E_- - 2E_0)/2 \approx q h^2$,
where $h$ is the magnetic field strength and $q=(g_I + g_J)^2 \mu_B^2/16 E_{\rm HFS}$,
$g_J$ and $g_I$ are the gyromagnetic ratios of the electron and the nucleus, $\mu_B$ is the Bohr magneton,
$E_{\rm HFS}$ is the hyperfine energy splitting at zero magnetic field. Finally, the spin-dependent Hamiltonian (\ref{EA}) becomes
\begin{equation}
H_{\rm A} = \int d^3r \, \left[ qh^2(n_+ + n_-) +\frac{c_2}{2} :{\bf F}^2: \right] \, ,
\end{equation}
where $F_z^2=n_+ - n_-$ and $F_{\perp}^2=2|\psi_+\psi_0^\dag + \psi_0\psi_-^\dag|^2$ are the square of magnetization density and the square of transverse spin density, respectively.
In spinor condensates realized in laboratories, $a_0$ and $a_2$ scattering lengths have similar magnitude. The spin-dependent interaction coefficient $c_2$ is, therefore, 
much smaller than its spin-independent counterpart $c_0$. For $^{23}$Na condensate their ratio is about 1:30 and is positive (antiferromagnetic order), while for $^{87}$Rb condensate it is 1:220 and is negative (ferromagnetic order).

By comparing the kinetic energy with the interaction energy, we can define the healing length $\xi=2\pi\hbar/\sqrt{2 m c_0 n}$ 
and the spin healing length $\xi_s=2\pi\hbar/\sqrt{2 m c_2 n}$. These quantities give the length scales of spatial variations in
the condensate profile induced by the spin-independent or spin-dependent interactions. Here we consider system sizes smaller than the spin healing length
in order to avoid a domain formation. A good basis for such a homogeneous system is the plane wave basis.

\section{The ideal gas}
\label{sec:idealgas}

We consider a uniform gas of non-interacting atoms ($c_0=c_2=0$) with hyperfine spin $F=1$ in a homogeneous magnetic field $h$ within the grand canonical ensemble,
taking into account the quadratic Zeeman effect. The effective Hamiltonian of the system is
\begin{equation}
H_{eff}=\sum_{m_F=1,0,-1}\sum_{{\bf k}} \left( \epsilon_{\bf k} + m_F^2 \, qh^2 \right) n_{{\bf k}, m_F} - \mu N -\eta M \, ,
\end{equation}
with
\begin{eqnarray}
N&=&N_{+}+N_0+N_-=\sum_{m_F} \sum_{{\bf k}} n_{{\bf k}, m_F} \, ,\\
M&=&N_{+}-N_-= \sum_{m_F} \sum_{{\bf k}} m_F n_{{\bf k}, m_F}\, .
\end{eqnarray}
Here ${\bf k}=2 \pi/L (n_x,n_y,n_z)$, $L$ is the system size and $n_l=0, \pm 1, \pm 2 \dots$ are integers, $n_{{\bf k}, m_F}$ are occupation numbers of atoms of energy $\epsilon_{\bf k}=\hbar^2{\bf k}^2/2m$. The chemical potential $\mu$ and the linear Zeeman shift $\eta$ are Lagrange multipliers enforcing the desired total atom number $N$ and the magnetization $M$ respectively. $N_{m_F}$ is the number of atoms in the $m_F$th component. We consider a positive magnetization $M \ge 0$ and a positive Zeeman energy shift $qh^2>0$.

The non-zero magnetic field removes the degeneracy of energy spectra:
\begin{eqnarray}
E_{{\bf k},+}&=& \epsilon_{\bf k} -\mu - \eta +qh^2,\\
E_{{\bf k},0}&=& \epsilon_{\bf k} -\mu,\\
E_{{\bf k},-}&=& \epsilon_{\bf k} -\mu + \eta +qh^2.
\end{eqnarray}
The ratio between $\eta$ and $qh^2$ determines the order of energy levels. The lowest energy level is $E_+$ for $qh^2 \le \eta$, or $E_0$ for $qh^2 \ge \eta$. In addition, two effects determine the state of the system: (i) the occupation number imbalance enforced by the fixed magnetization $N_+\ge N_-$, and (ii) the ground state energy level ($E_0$ or $E_+$) controlled by the magnetic field.

Since the Hamiltonian is diagonal, we may calculate the grand canonical partition function
\begin{equation}
\Xi=\sum_{m_F, n_{{\bf k}, m_F}} {\rm e} ^{-\beta E_{{\bf k},m_F} n_{{\bf k},m_F}} \, ,
\end{equation}
where $\beta=1/k_B T$, $T$ is the temperature and $k_B$ is the Boltzmann constant. The ensemble average of the occupation number $n_{{\bf k}, m_F}$ is
\begin{equation}
n_{{\bf k}, m_F}=-\frac{1}{\beta} \frac{\partial {\rm ln} \Xi}{\partial E_{{\bf k}, m_F}} \, ,
\end{equation}
which gives
\begin{equation}
n_{{\bf k}, m_F}=\frac{z_{m_F} {\rm e}^{-\beta \epsilon_{\bf k}}} {1-z_{m_F} {\rm e}^{-\beta \epsilon_{\bf k}}} \, 
\end{equation}
with effective fugacities
\begin{eqnarray}
z_+&=& e^{\beta(\mu+\eta-qh^2)} \, ,\\
z_0&=& e^{\beta\mu}\, ,\\
z_-&=& e^{\beta(\mu-\eta-qh^2)} \, .
\end{eqnarray}
In the thermodynamic limit, keeping only dominant terms $O(N)$, expressed in terms of fugacities, the number of atoms in the lowest energy level of each $m_F$ component is
\begin{equation}
N_{m_F}^c=\frac{z_{m_F}}{1-z_{m_F}} \, ,
\label{eq:condN+}
\end{equation}
while the number of thermal atoms in each $m_F$ component is
\begin{equation}
N_{m_F}^T=\left(\frac{L}{\lambda_{dB}}\right)^3g_{\frac{3}{2}}(z_{m_F}) \, ,
\end{equation}
where $\lambda_{dB}=h/\sqrt{2 \pi m k_B T}$ is the de Broglie wave length, and $g_j(x)=\sum_{n=1}^{+\infty} x^n/n^j$ is the Bose function.

\subsection{Transition temperatures and condensate fractions}

\subsubsection{For $qh^2 \le \eta$}
\label{sec:qh2leeta}

\emph{The first phase transition} occurs for $z_+\to 1$ (or $\mu \to qh^2 -\eta$) when the $m_F=1$ component starts condensing. 
That $N_+^c \gg 1$ can be seen from (\ref{eq:condN+}). The number of thermal atoms is then
\begin{eqnarray}
N_+^T& = & \left(\frac{L}{\lambda_{dB}}\right)^3g_{\frac{3}{2}}(1) \, ,
\label{eq:highM+} \\
N_0^T&=& \left(\frac{L}{\lambda_{dB}}\right)^3g_{\frac{3}{2}}(e^{\beta qh^2}z_{\eta}) \, ,
\label{eq:highM0} \\
N_-^T&=&\left(\frac{L}{\lambda_{dB}}\right)^3g_{\frac{3}{2}}(z_{\eta}^2) \, ,
\label{eq:highM-}
\end{eqnarray}
with $z_{\eta}\equiv e^{-\beta\eta}$. The first critical temperature $T_{c1}$ can be obtained from the following equations:
\begin{eqnarray}
N &=& \left(\frac{L}{\lambda_{dB}(T_{c1})}\right)^3 F^+_{\frac{3}{2}}\left(T_{c1}, z_{\eta c1} \right) \, ,\\
M &=& \left(\frac{L}{\lambda_{dB}(T_{c1})}\right)^3 \left( g_{\frac{3}{2}}(1)-g_{\frac{3}{2}} \left( z_{\eta c1}^2 \right) \right) \, ,
\end{eqnarray}
where we have introduced the notation $z_{\eta c1}\equiv z_{\eta}(T_{c1})$ and
\begin{equation}
F^+_{\frac{3}{2}}(T,z_{\eta})\equiv g_{\frac{3}{2}}(1)+g_{\frac{3}{2}}(e^{\beta qh^2}z_{\eta})+g_{\frac{3}{2}}(z_{\eta}^2) \, .
\end{equation}

Below $T_{c1}$, only the $m_F=1$ component condenses. It is justified to assume $N_+^c \simeq N^c$. Then
relation $N^c=N-\sum_{\sigma}N_\sigma^T$ defines the condensate fraction of the $m_F=1$ component
\begin{equation}
\frac{N_+^c}{N} \simeq1-\left(\frac{T}{T_{c1}}\right)^{\frac{3}{2}}\frac{F^+_{\frac{3}{2}}(T,z_{\eta})}{F^+_{\frac{3}{2}}(T_{c1},z_{\eta c1})} \, .
\end{equation}

\emph{The second phase transition} occurs for $z_{\eta} \to e^{-\beta qh^2}$ ($\eta \to qh^2$) when $N_0^c \gg 1$ and
$N_-^c\to {\rm e}^{-2 \beta qh^2}/(1-{\rm e}^{-2 \beta qh^2})$. In this regime, $T<T_{c2}$, thermal populations are
\begin{eqnarray}
N_+^T&=&\left(\frac{L}{\lambda_{dB}}\right)^3g_{\frac{3}{2}}(1) \, , \\
N_0^T&=&\left(\frac{L}{\lambda_{dB}}\right)^3g_{\frac{3}{2}}(1) \, , \\
N_-^T&=&\left(\frac{L}{\lambda_{dB}}\right)^3g_{\frac{3}{2}}(e^{-2\beta qh^2})\, .
\end{eqnarray}
The second transition temperature $T_{c2}$ can be obtained using the difference between the total atom number $N$ and the magnetization $M$. For temperatures $T\in[T_{c2},T_{c1}]$, in the absence of condensates in $m_F=0,-1$ components, the difference is $N-M \simeq 2N_-^T+N_0^T$. The second transition temperature expressed in terms of Bose functions present in equations (\ref{eq:highM0}) and (\ref{eq:highM-}) is
\begin{equation}
k_B T_{c2} = \frac{2 \pi \hbar^2}{m L^2}   \left[\frac{N-M}{G_{\frac{3}{2}}(T_{c2})} \right]^{\frac{2}{3}} \, ,
\end{equation}
where
\begin{equation}
G_{\frac{3}{2}}(T) \equiv g_{\frac{3}{2}}(1)+2g_{\frac{3}{2}}\left({\rm e}^{-2\beta qh^2}\right) \, .
\end{equation}

Below $T_{c2}$, the Bose-Einstein condensate can be formed in all components and condensate fractions satisfy the set of equations:
\begin{equation}
    \begin{array}{rl}
      N^c=&M^{c}(T)+(N-M)\left[1-\left(\frac{T}{T_{c2}}\right)^{\frac{3}{2}} \frac{G_{\frac{3}{2}}(T)}{G_{\frac{3}{2}}(T_{c2})}\right]\, ,\\
      N_+^c-&N_-^c=M^{c}(T) \, ,\\
      \frac{2}{N_0^c}=&\frac{1}{N_+^c}+\frac{e^{-2\beta qh^2}}{N_-^c}-2\sinh(\beta qh^2)e^{-\beta qh^2} \, .
    \end{array}
\label{eq:cf1-3}
\end{equation}
Here we have introduced the condensate part of the magnetization $M^{c} \equiv M- M^{T}$ and the thermal part of magnetization
\begin{equation}
M^{T}(T) \equiv  \left(\frac{L}{\lambda_{dB}} \right)^3 \left[ g_{\frac{3}{2}}(1) - g_{\frac{3}{2}}\left({\rm e}^{-2\beta qh^2}\right) \right] \, .
\end{equation}
A derivation of eqs. (\ref{eq:cf1-3}) is included in Appendix \ref{app:equation for condfrac}. An analytical solution of eqs. (\ref{eq:cf1-3}) is presented in Appendix \ref{app:solution for condfrac}. We have checked the validity of the analytical solution against the self-consistent numerical result.

\subsubsection{For $qh^2 \ge \eta$}
\label{sec:qh2geeta}
First, one should notice that this case does not exist in the absence of an external magnetic field, since $\eta$ can take positive values for $M>0$. That is a new area of the phase diagram, which appears due to the quadratic Zeeman effect.

\emph{The first phase transition.} This time, the $m_F=0$ component undergoes condensation first, which means that $z_0 \to 1$ (or $\mu \to 0$) and $N_0^c\gg 1$. One obtains new expressions for $N_+^T$, $N_0^T$ and $N_-^T$, which hold under the critical temperature $T_{c1}$:
\begin{eqnarray}
N_+^T&=&\left(\frac{L}{\lambda_{dB}}\right)^3g_{\frac{3}{2}}(e^{-\beta qh^2}{z_{\eta}}^{-1}) \, ,
\label{eq:lowM+}\\
N_0^T&=&\left(\frac{L}{\lambda_{dB}}\right)^3g_{\frac{3}{2}}(1) \, ,
\label{eq:lowM0}\\
N_-^T&=&\left(\frac{L}{\lambda_{dB}}\right)^3g_{\frac{3}{2}}(e^{-\beta qh^2}z_{\eta}) \, .
\label{eq:lowM-}
\end{eqnarray}
The first critical temperature $T_{c1}$ and the fugacity at the critical point $z_{\eta c1}$ can be obtained from the following equations:
\begin{eqnarray}
N&=&\left( \frac{L}{\lambda_{dB}(T_{c1})} \right)^3 F^0_{\frac{3}{2}}\left(T_{c1},z_{\eta c1} \right) \, , \\
M&=& \left( \frac{L}{\lambda_{dB}(T_{c1})} \right)^3 \left[g_{\frac{3}{2}} ( {z_{\eta c1}}^{-1} e^{\frac{-qh^2}{kT_{c1}}} ) -
 g_{\frac{3}{2}} ({z_{\eta c1}} e^{\frac{-qh^2}{kT_{c1}}} ) \right]  \, ,\nonumber\\
\end{eqnarray}
where we have introduced
\begin{equation}
F^0_{\frac{3}{2}}(T,z_{\eta})\equiv g_{\frac{3}{2}}({z_{\eta}}^{-1}e^{-\beta qh^2})+g_{\frac{3}{2}}(1)+g_{\frac{3}{2}}({z_{\eta }}e^{-\beta qh^2}) \, .
\end{equation}
At the critical the point the fugacity is smaller than one ($z_{\eta}(T_{c1})<1$) since $M \ge 0$ and $g_{\frac{3}{2}}$ is an increasing function of its argument and takes positive values.

Assuming that $N_0^c \simeq N^c$ for $T\in[T_{c2},T_{c1}]$, once again the relation $N^c=N-\sum_{m_F}N_{m_F}^T$ defines the condensate fraction in the $m_F=0$ component
\begin{equation}
\frac{N_0^c}{N} \simeq1-\left(\frac{T}{T_{c1}}\right)^{\frac{3}{2}}\frac{F^0_{\frac{3}{2}}(T,z_{\eta})}{F^0_{\frac{3}{2}}(T_{c1},z_{\eta c1})} \, .
\end{equation}

\emph{The second phase transition}. One expects $z_{\eta} \sim e^{-\beta qh^2}$, implying that the $m_F=1$ component starts condensing: $N_+^c\gg1$ and again $N_-^c\to {\rm e}^{-2 \beta qh^2}/(1-{\rm e}^{-2 \beta qh^2})$. Nevertheless, in this regime we should define $T_{c2}$ in the other way. Neither $N$ nor $N-M$ can be used anymore, since they involve $N_c^0/N$ which is now unknown in the intermediate region of temperatures. The only solution is to use the magnetization $M$, and define $T_{c2}$ as the temperature for which $N_+^c\simeq N_-^c<<N$, that is
\begin{equation}
k_B T_{c2} = \frac{2 \pi \hbar^2}{m L^2}\left(\frac{M}{g_{\frac{3}{2}}(1)-g_{\frac{3}{2}}(e^{\frac{-2qh^2}{kT_{c2}}})}\right)^{\frac{2}{3}} \, ,
\end{equation}
what is equivalent to $M=M^T(T_{c2})$. This choice is justified in the thermodynamic limit when $N>>{\rm e}^{-2 \beta qh^2}/(1-{\rm e}^{-2 \beta qh^2})$, and mathematically within our equations for $\beta qh^2\gg 1$ and any $N$.

Below $T_{c2}$, condensate fractions satisfy the set of equations:
\begin{equation}
    \begin{array}{rl}
      N^c=&N-\left( \frac{L}{\lambda_{dB}} \right)^3 \left[2 g_{\frac{3}{2}}(1) + g_{\frac{3}{2}}({\rm e}^{-2\beta qh^2}) \right] \, , \\
      N_+^c-&N_-^c=M^c(T) \, ,\\
      \frac{2}{N_0^c}=&\frac{1}{N_+^c}+\frac{e^{-2\beta qh^2}}{N_-^c}-2\sinh(\beta qh^2)e^{-\beta qh^2} \, .
    \end{array}
\label{eq:highqh2cf1-3}
\end{equation}
Different from (\ref{eq:cf1-3}) is the first equation of (\ref{eq:highqh2cf1-3}) only.

\subsection{Phase diagram}

\begin{figure*}[!]
\centerline{\includegraphics[width=4.5cm, keepaspectratio, angle=-90]{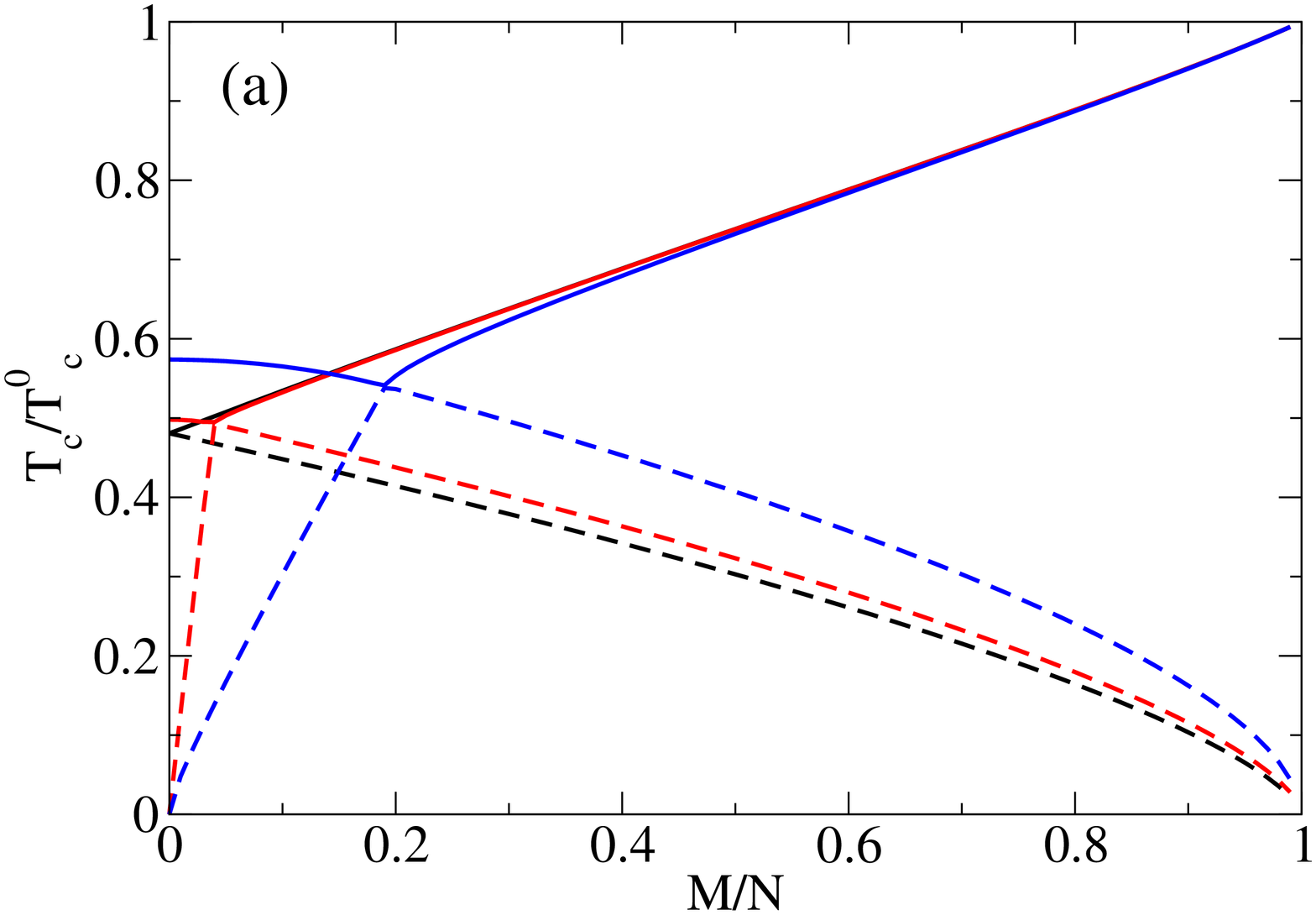}
\includegraphics[width=4.5cm, keepaspectratio, angle=-90]{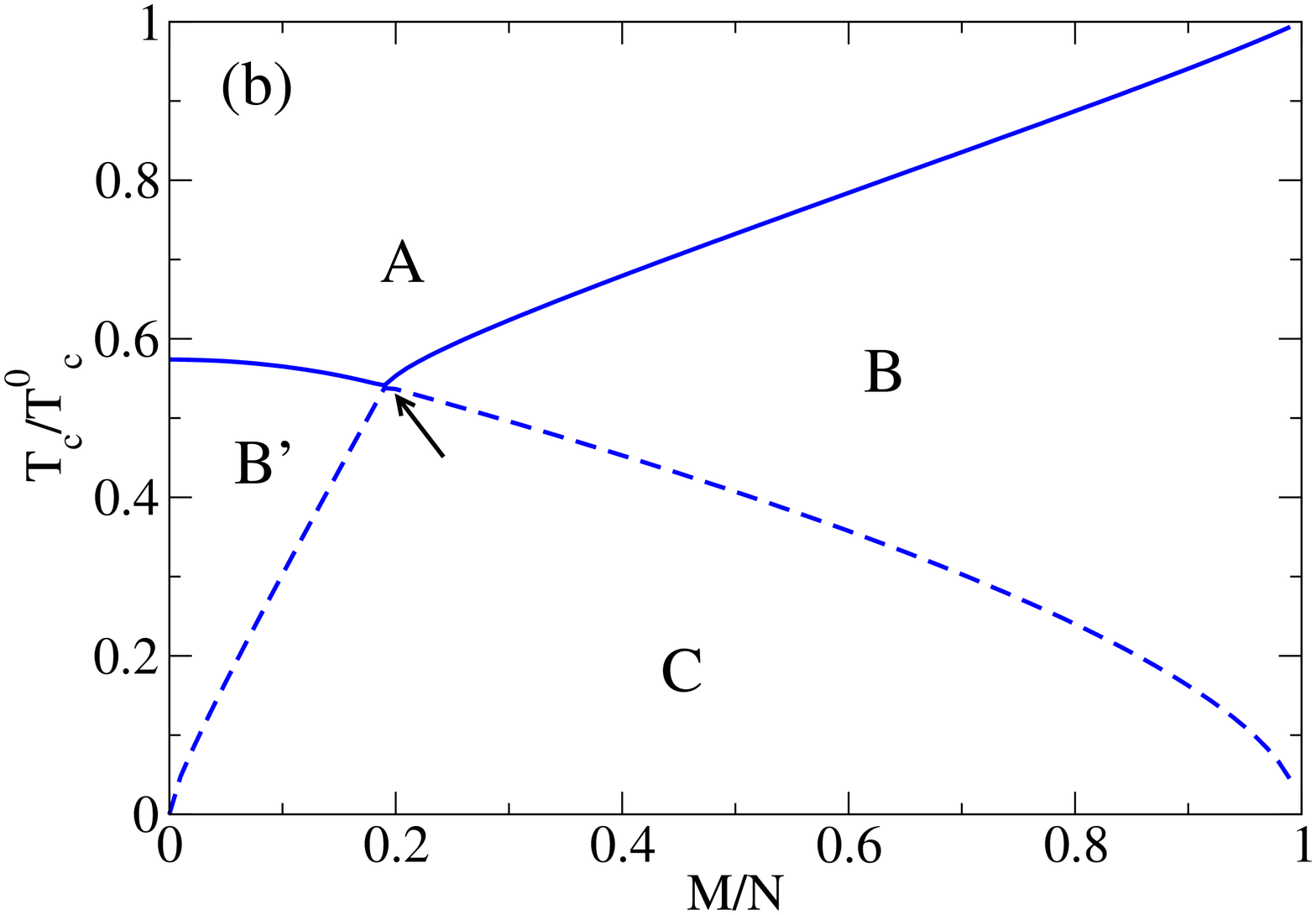}
\includegraphics[width=4.5cm, keepaspectratio, angle=-90]{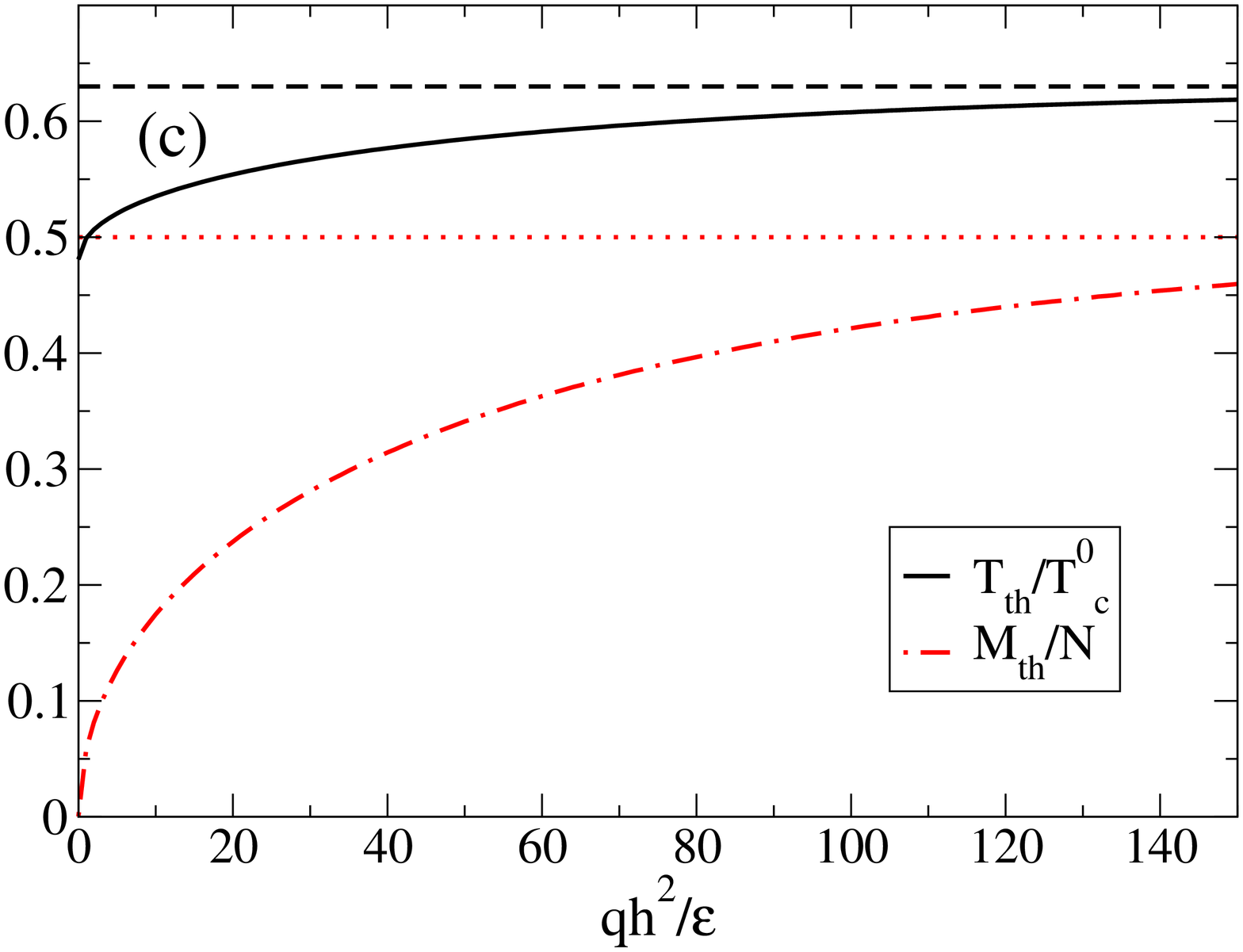}}
\caption{(Color online) (a) and (b) show the phase diagram of the critical temperatures. $T_{c1}$ is marked by solid lines and $T_{c2}$ is marked by dashed lines. 
In (a) black lines are for $qh^2=0$, red lines are for $qh^2=0.5 \hbar^2/mL^2$ and blue lines are for $qh^2=12.5 \hbar^2/mL^2$. 
$T^0_c=(2 \pi \hbar^2/mL^2)(N/\zeta(3/2))^{2/3}$ is the critical temperature for one component condensate in the box potential. 
In (b) the same for $qh^2=12.5 \hbar^2/2 mL^2$. The arrow indicates the threshold at the critical temperatures intersection point. 
Particular parts of the diagram are "$A$"-thermal atoms (no condensate), "$B$"-condensate in the component $m_F=1$, 
"$B'$"-condensate in the component $m_F=0$ only, "$C$"-condensate possible in all components.
(c) The threshold temperature $T_{th}/T^0_c$ (solid line) and the threshold magnetization $M_{th}/N$ (dash-dotted line) 
at the critical temperatures intersection point as a function of the quadratic Zeemann energy shift $qh^2/\varepsilon$ with $\varepsilon=\hbar^2/2 mL^2$. 
The asymptotic values of the threshold critical temperature $T^{\infty}_{th}/T^{0}_{th} = 2^{-2/3}$ and the threshold magnetization $M^{\infty}_{th}=1/2$ 
are marked by dashed and dotted lines respectively. Here, the total number of atoms is $N=10^4$.}
\label{phasediagrammh}
\end{figure*}

The non-zero magnetic field changes dramatically the phase diagram of the critical temperatures, which is shown in Fig. \ref{phasediagrammh}. The phase diagram consists of four phases $"A",\, "B", \, "B'", \, "C"$ separated by the two critical temperatures $T_{c1}$ and $T_{c2}$. Depending on the value of the temperature, the system can be: $"A"$- a non-degenerate thermal gas, $"B"$- condensate in the component $m_F=1$ or $"B'"$-condensate in the $m_F=0$ and thermal atoms in other components, $"C"$- a condensate in $m_F=0$ and $m_F=1$ while in the $m_F=-1$ is a gas with non-negligible fraction of atoms in the lowest energy level for $\beta qh^2 \ll 1$, or with negligible fraction of atoms in the lowest energy level for $\beta qh^2 \gg 1$. A possible destination is controlled by the magnetization, with a special role of the threshold magnetization at the critical temperatures intersection point $M_{th}\equiv M(T=T_{c1}=T_{c2})$. If $M<M_{th}$ then to obtain particular quantities, one should use expressions from subsection \ref{sec:qh2geeta}, in the opposite case ($M>M_{th}$) from subsection \ref{sec:qh2leeta}. The procedure to obtain numerical values for the critical temperatures is explained in Appendix \ref{diagrammephaseh}.

Analytical expressions for the threshold critical temperature $T_{th}$ and the threshold magnetization $M_{th}$ are
\begin{eqnarray}
\left( \frac{T_{th}}{C} \right)^{3/2}& = & \frac{N}{2 g_{3/2}(1) + g_{3/2}\left( e^{-2 qh^2/k_B T_{th}}\right)} \, , \label{eq:Tth} \\
\frac{M_{th}}{N}&=& \frac{3 g_{3/2}(1)}{N} \left( \frac{T_{th}}{C} \right)^{3/2} -1 \, , \label{eq:Mth}
\end{eqnarray}
where $C=h^2/2 \pi m L^2 k_B$. The above expressions are obtained by comparing critical temperatures $T_{c1}$ and $T_{c2}$ for both $qh^2>\eta$ and $qh^2<\eta$. In fig.~\ref{phasediagrammh}c we show the threshold critical temperature (\ref{eq:Tth}), the threshold magnetization (\ref{eq:Mth}) and their asymptotic values for $\beta qh^2\to \infty$, they are $T^{\infty}_{th}/T^{0}_{th} \to 2^{-2/3}$ and $M^{\infty}_{th}/N\to 1/2$ respectively.

\subsection{Condensed fractions}

\begin{figure*}
\centerline{\includegraphics[width=4.5cm, keepaspectratio, angle=-90]{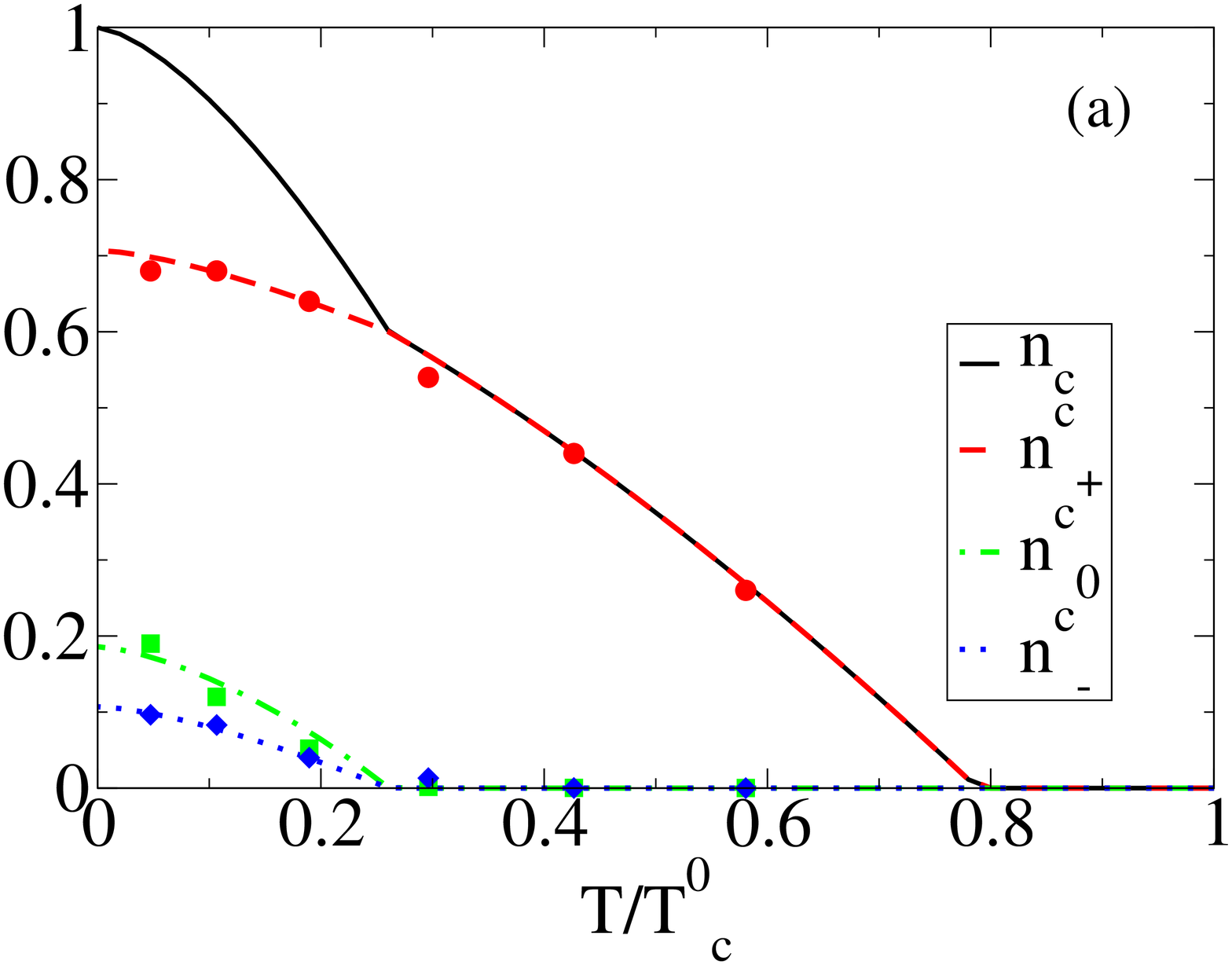}
\includegraphics[width=4.5cm, keepaspectratio, angle=-90]{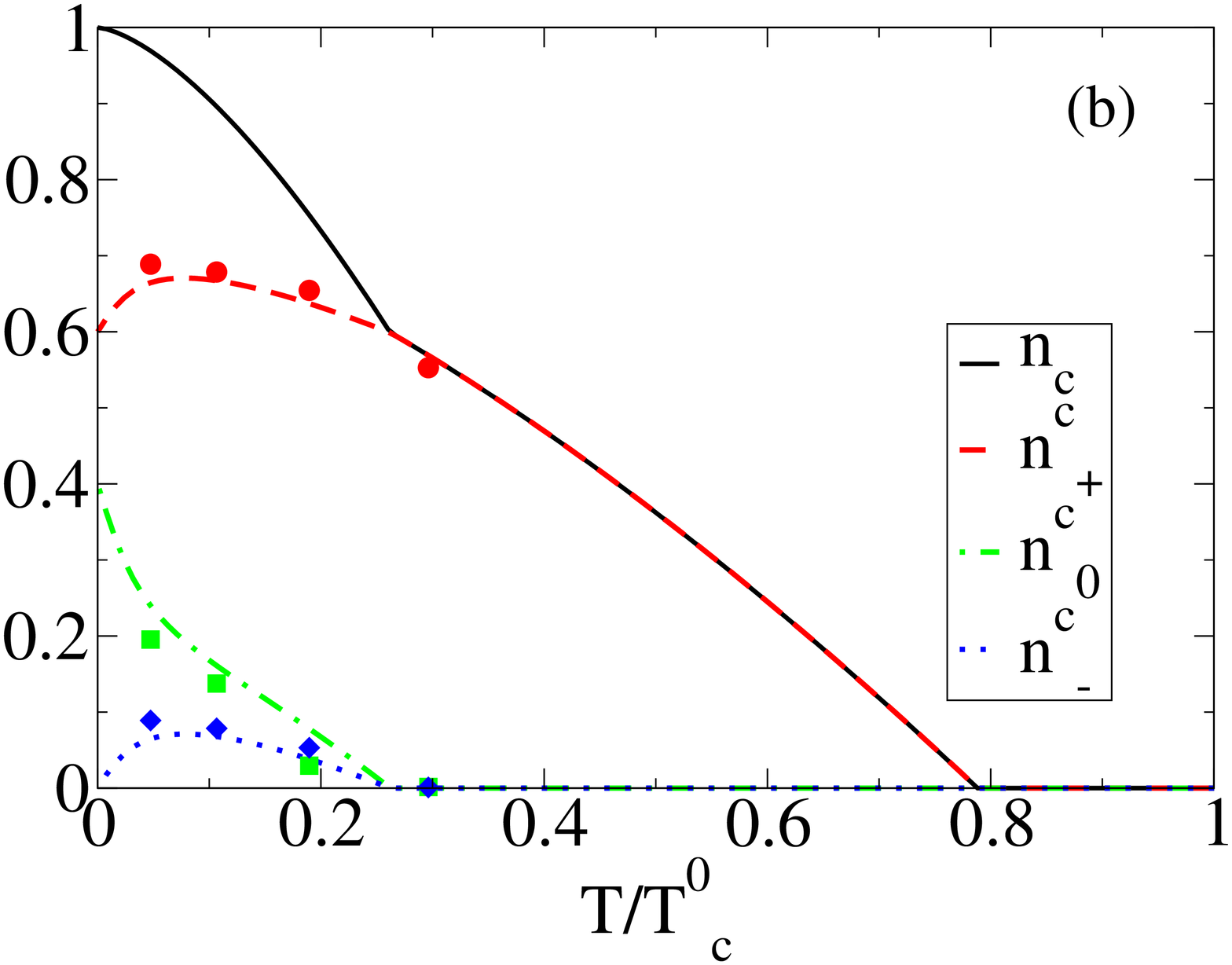}
\includegraphics[width=4.5cm, keepaspectratio, angle=-90]{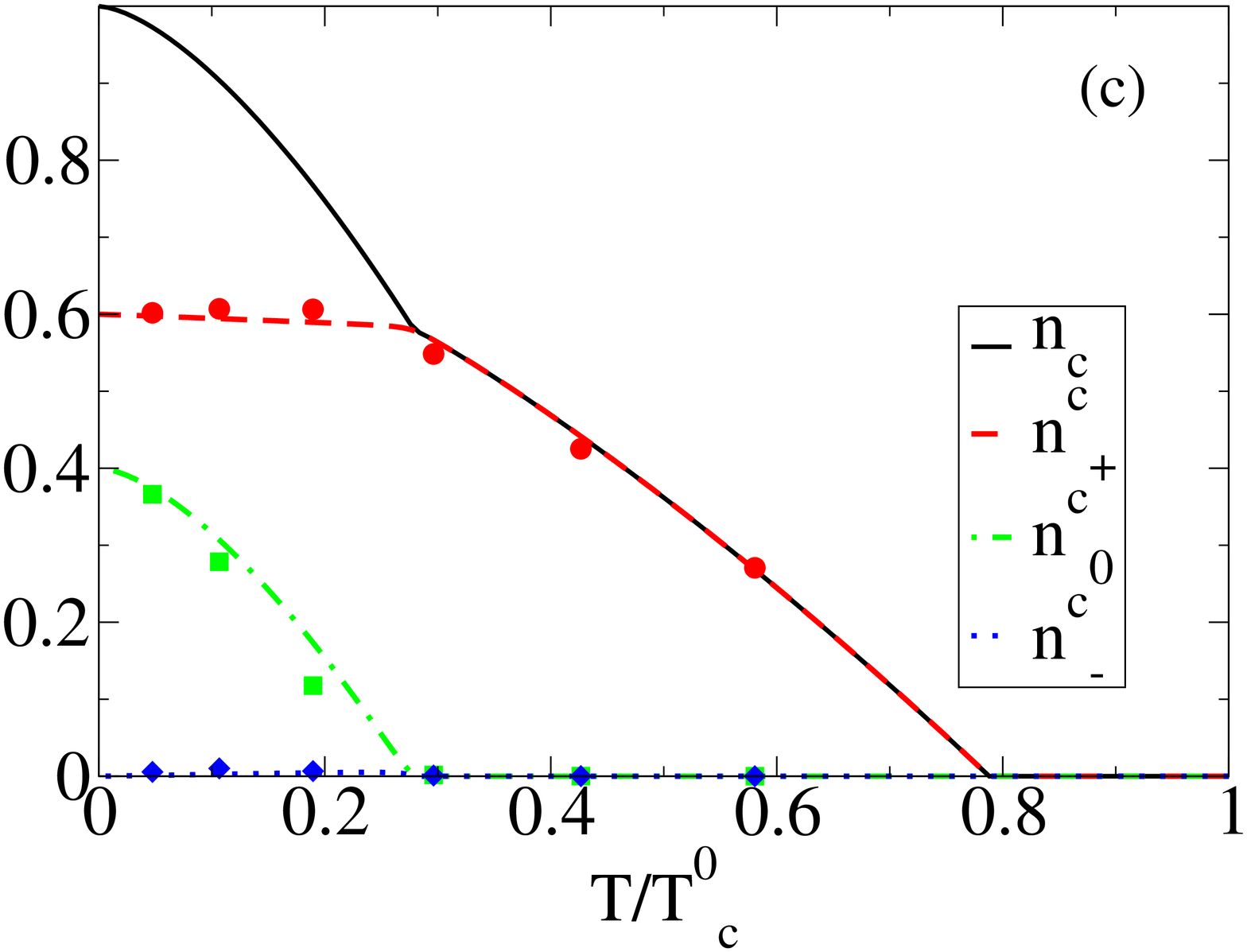}}
\caption{(Color online) Condensate fractions for $M>M_{th}$, with $N=10^4$ and $M=6\times 10^3$; (a) $qh^2=0$, (b) $qh^2=0.1\, \hbar^2/mL^2$ and (c) $qh^2=\hbar^2/mL^2$. 
$n_c=N^c/N$ is the total condensate fraction (solid black line), $n_+^c$ is the condensate fraction in $m_F=1$ component (dashed red line), and $n_0^c$, $n_-^c$ in $m_F=0$ (dot-dashed green line), $m_F=-1$ (dotted blue line) respectively. Lines are the solution of equations (\ref{eq:cf1-3}) while points are results of Monte Carlo simulations.}
\label{condensatefrac1}
\end{figure*}
\begin{figure*}
\centerline{\includegraphics[width=4.5cm, keepaspectratio, angle=-90]{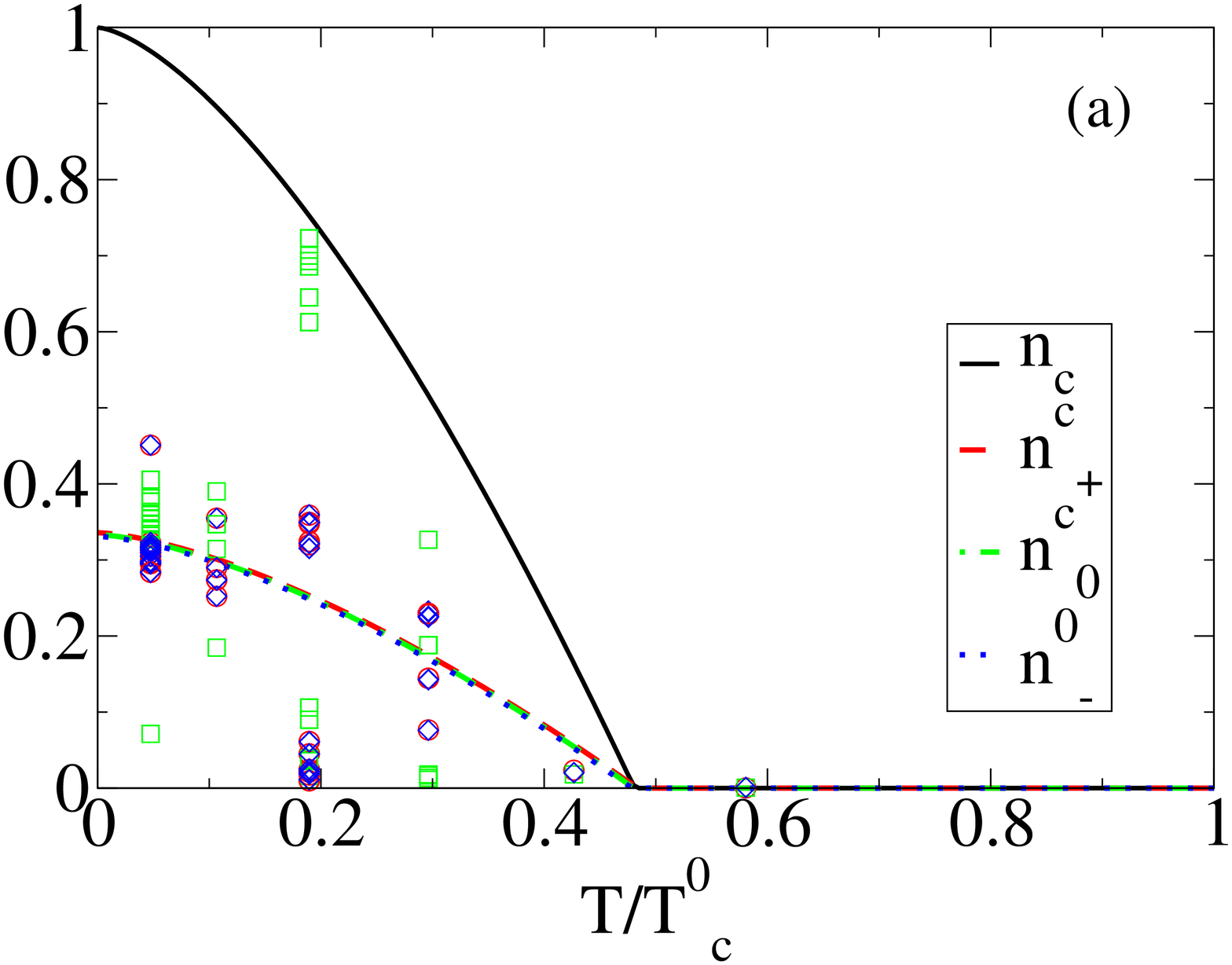}
\includegraphics[width=4.5cm, keepaspectratio, angle=-90]{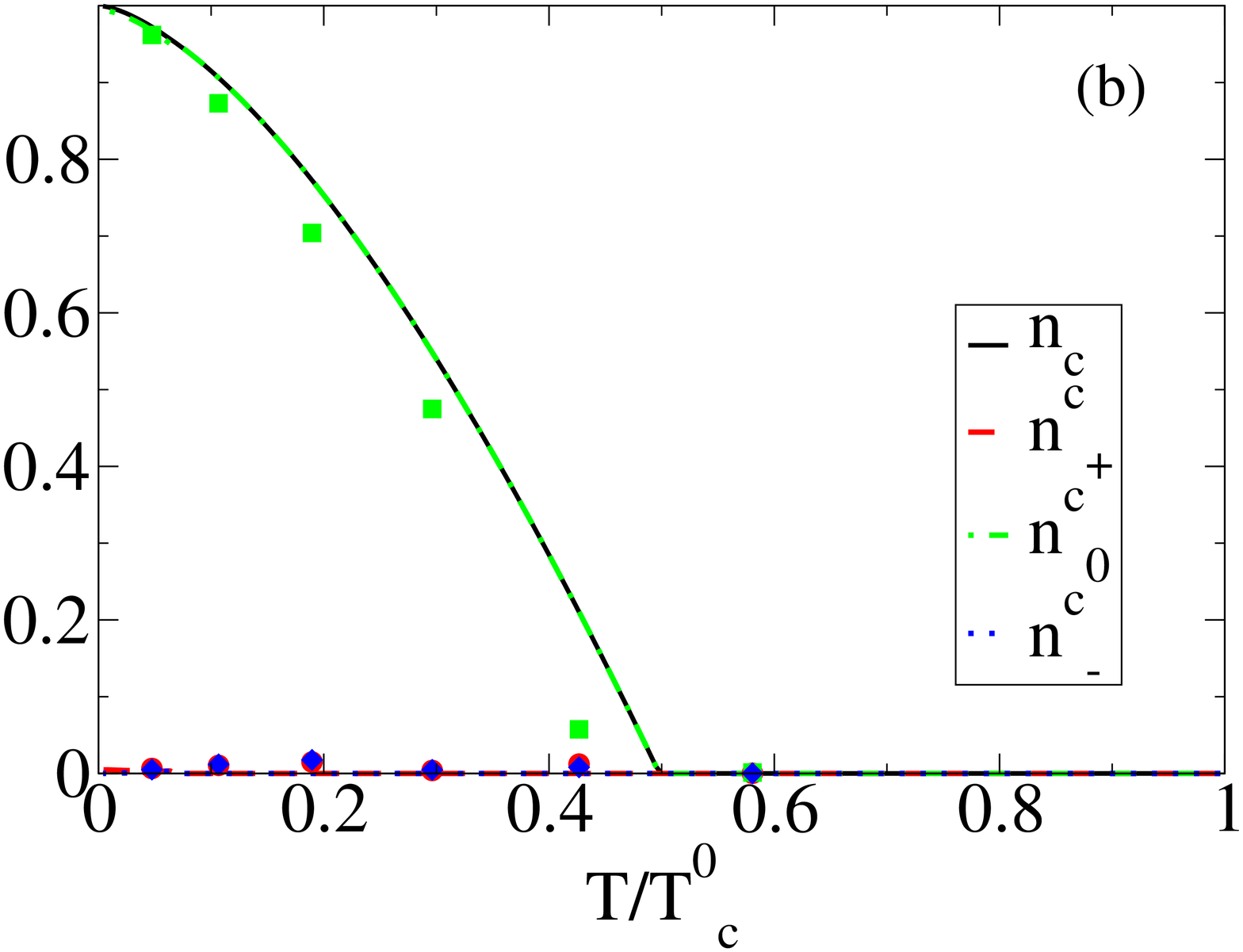}}
\caption{(Color online) Condensate fractions for $M<M_{th}$, with $N=10^4$ and $M=50$; (a) $qh^2=0$ and (b) $qh^2=\hbar^2/mL^2$.  
$n_c=N^c/N$ is the total condensate fraction (solid black line), 
$n_+^c$ is the condensate fraction in the $m_F=1$ component (dashed red line), 
and $n_0^c$, $n_-^c$ in $m_F=0$ (dot-dashed green line), $m_F=-1$ (dotted blue line) respectively. Lines are solution of equations (\ref{eq:highqh2cf1-3}) while points are results of Monte Carlo simulations. Particular points in (a) correspond to averaging over different representations of an ensemble and show strong fluctuations of condensate fractions in the regime of zero magnetic field and almost zero magnetization.}
\label{condensatefrac2}
\end{figure*}

Condensate fractions, solutions of equations (\ref{eq:cf1-3}) for $M>M_{th}$ and solutions of equations (\ref{eq:highqh2cf1-3}) for $M<M_{th}$, are presented in fig. \ref{condensatefrac1} and \ref{condensatefrac2} respectively and they are marked by lines. Points are results of the Metropolis algorithm adapted to the model (more details concerning the algorithm can be found in section \ref{app:MonteCarlo}).

Figure \ref{condensatefrac1} is for values of the magnetic field in the area $M>M_{th}$ where $qh^2 \le \eta$. These graphs show that modifications of condensed fractions occur mainly for low magnetic fields. The effect of the non-zero magnetic field is the most visible on the condensate fraction in the $m_F=-1$ component. 
Notice, at zero magnetic field the fraction of the condensate in the $m_F=-1$ component decreases simply with the temperature, see fig.~\ref{condensatefrac1}a.
In the transient magnetic field regime, the condensed fraction in the $m_F=-1$ component increases from zero, reaches a maximum and then decreases to zero at the second critical temperature, see fig.\ref{condensatefrac1}b. The condensed fraction in the $m_F=-1$ component decreases quickly and can be neglected in the high magnetic field regime, see fig.~\ref{condensatefrac1}c. 
Condensed fractions are linked together, when the condensed fraction of the $m_F=-1$ component disappears, in the meantime the condensed fraction in the $m_F=1$ component decreases and the condensate fraction in the $m_F=0$ increases. Nevertheless, the slope-breaking that occurs at $T_{c2}$, already present when $h=0$, is still neat. The fugacity varies dramatically near the zero temperature for small values of magnetic fields, what explains sharp variations of the condensed fractions in fig.~\ref{condensatefrac1}b.

Figure \ref{condensatefrac2} is for the magnetization $M<M_{th}$ where $qh^2 \ge \eta$. The value of magnetization is $M=50$ and is very small as compared to $N=10^4$, therefore the difference between $N_+^c$ and $N_-^c$ is not visible. Notice, strong fluctuations of the condensate fractions for zero magnetic field that are results of Monte Carlo simulations, see different points in fig.~\ref{condensatefrac2}a. Indeed, for zero magnetization and zero magnetic field the ground state of the ideal gas is strongly degenerate~\cite{Zhang}, what gives rise to strong fluctuations of condensate fractions. The non-zero magnetic field reduces degeneracy and hence reduces fluctuations of condensate fractions in fig.~\ref{condensatefrac2}b.

\section{The interacting gas}
\label{sec:With interactions}

The ground state of a spin-1 Bose gas in the presence of ferromagnetic and antiferromagnetic interactions was widely studied within the single-mode aproximation~\cite{GS_theory} and beyond~\cite{Matuszewski_AF}, and was investigated in experiments for antiferromagnetic condensates~\cite{GS_experiment}. The structure of the ground state is quite complex and depends not only on the magnetization and magnetic field but also on the relative phase between components of the Bose gas. It consists of a polar, nematic or magnetic state, two component or three component solutions with phase and anti-phase matching for ferromagnetic and antiferromagnetic interactions respectively. We are aware of the temperature dependence of such structures, in particular the boundaries between different phases. 

The non-zero temperature introduces a multi-mode structure, therefore we describe the system within the classical fields approximation that takes into account thermal populations and interactions among many modes. Indeed, classical fields and stochastic methods \cite{SGPE} as well as Hartree-Fock or Hartree-Fock-Popov approximations \cite{other_meth} were applied for spinor condensates at non-zero temperature but for free magnetization. Among all of finite temperature methods that are used for single-component condensates, the classical fields-like are not perturbative and thus contain all nonlinear terms that are present in the Hamiltonian. It makes them very suitable for study thermodynamics in the whole temperature range, what is not a case for methods based on the Bogoliubov approximation. 

Below we just briefly remind the main concept of the classical field approach, more details concerning the foundations of the approximation can be found in~\cite{CFM}.

\subsection{The classical fields approximation}

The classical fields approach consists of (i) replacement of the creation and annihilation operators by complex amplitudes,
(ii) restriction of the summation over modes to a finite number extended all the way to the momentum cut-off ${\bf K}_{max}$.

The field operator is replaced by a classical field (complex function) of well-defined number of momenta modes:
\begin{equation}
\psi_j({\bf r})=\sum_{{\bf k}\le {{\bf K}_{max}}} a_j({\bf k}) \, .
\end{equation}
The energy $E_{\psi}$ of such a classical field is given by discretization of the hamiltonian $H=H_0 + H_A$, eqs.~(\ref{En}) and (\ref{EA}). The total number of atoms is
\begin{equation}
N=\sum_{j} \sum_{\bf k\le {{\bf K}_{max}}} |a_j({\bf k})|^2
\end{equation}
and the magnetization is
\begin{equation}
M=\sum_{{\bf k\le {{\bf K}_{max}}}} \left( |a_+({\bf k})|^2 - |a_-({\bf k})|^2 \right).
\end{equation}

Various observables have a more or less pronounced dependence on the cut-off. 
Here we choose the cut-off momentum such, that in the thermodynamic limit the non-condensed density for a single component ideal Bose gas in degenerate regime is exactly reproduced by the classical field model~\cite{our_EPJST}. The condition gives $E_{{\bf K}_{max}}\simeq 2.695 k_BT$, where $E_{{\bf K}_{max}}=3\hbar^2 (\pi/L)^2/2 m$ is the maximal kinetic energy on the grid.

\subsection{The Metropolis algorithm for a spin-1 Bose gas with fixed magnetization}
\label{app:MonteCarlo}

We adapt the Metropolis scheme \cite{metropolis} to the system of classical fields as described in \cite{Optc}. The main idea of this Monte Carlo method is to generate a Markovian process of a random walk in phase space. All states of the system visited during this walk become members of the statistical ensemble and are used in ensemble averages. 

In order to obtain a statistical average of any observable $A$:
\begin{equation}
\bar{A}_j=\frac{1}{\mathcal N} \sum_{s=1}^{\mathcal N} \left< \psi_j^{(s)} | A | \psi_j^{(s)} \right> \,
\end{equation}
one should generate ${\mathcal N}$ copies of the classical fields $\psi_j^{(s)}$. 
A canonical average is obtained in the limit of ${\mathcal N}\to \infty$ provided the number of members of the ensemble with energy $E_{\psi}$ is proportional to the Boltzmann 
factor $e^{-E_{\psi}/k_BT}$.
This can be achieved in a random walk where a single step of the Markov process is defined as follows:
\begin{enumerate}
\item A set of amplitudes $ a_j^{(s)}({\bf k}) $ determines the state selected to be a member of the canonical ensemble at the $s$th step of the random walk.
The corresponding energy $E_{\psi}$ of the classical field is calculated according to (\ref{En}) and (\ref{EA}).
As the initial condition ($s=1$), any state that satisfies the condition of the fixed total number of atoms $N$ and the magnetization $M$ may be chosen as a member of the ensemble.
\item
A trial set of amplitudes $ \tilde{a}_j^{(s)}({\bf k}) $ is generated by a random disturbance of $ \tilde{a}_j^{(s)}({\bf k})=a_j^{(s)}({\bf k})  +  \delta_j^{(s)}({\bf k}) $ followed by normalization to account the total number of atoms. This way a trial classical field $\tilde{\psi}_j^{(s)}$ is obtained. The corresponding magnetization $\tilde{M}_s$, energy $\tilde{E}_{\tilde{\psi}}$, the energy difference $\Delta_s=E_{\psi} - \tilde{E}_{\tilde{\psi}}$, as well as the Boltzmann factor $p_s=e^{-\Delta_s/k_BT}$ are then calculated.
\item
If the magnetization $\tilde{M}_s$ of a trial set of amplitudes satisfy $|M-\tilde{M}_s|\le \delta M$ then $ \tilde{a}_j^{(s)}({\bf k}) $ can be considered as a new member of the ensemble.
\item
A new member of the Markov chain $ a_j^{(s+1)}({\bf k}) $ is selected according to the following prescription:
$(i)$ if $\Delta_s<0$ then the trial state becomes a new member of the ensemble $ a_j^{(s+1)}({\bf k}) = \tilde{a}_j^{(s)}({\bf k})$,
$(ii)$ if $\Delta_s>0$ then a random number $0<u<1$ is generated. If $u<p_s$ then the trial state becomes a new member of the ensemble. 
In the opposite case $u>p_s$, the "initial" state $ a_j^{(s)}({\bf k}) $ is once more included in the ensemble $ a_j^{(s+1)}({\bf k}) = a_j^{(s)}({\bf k})$ .
\end{enumerate}
The convergence of the procedure is the fastest when approximately every second trial state becomes a member of the ensemble. This factor depends on the assumed maximal value of displacements
$\delta_j^{(s)}({\bf k})$ which can be modified during the walk. The parameter $\delta M$ should be small enough to ensure almost constant magnetization $M$.
Note, some number of initial members of the ensemble should be ignored in order to avoid an influence of the arbitrarily selected initial state of the system.

\begin{figure}[]
\centerline{\includegraphics[width=3.6cm, keepaspectratio, angle=-90]{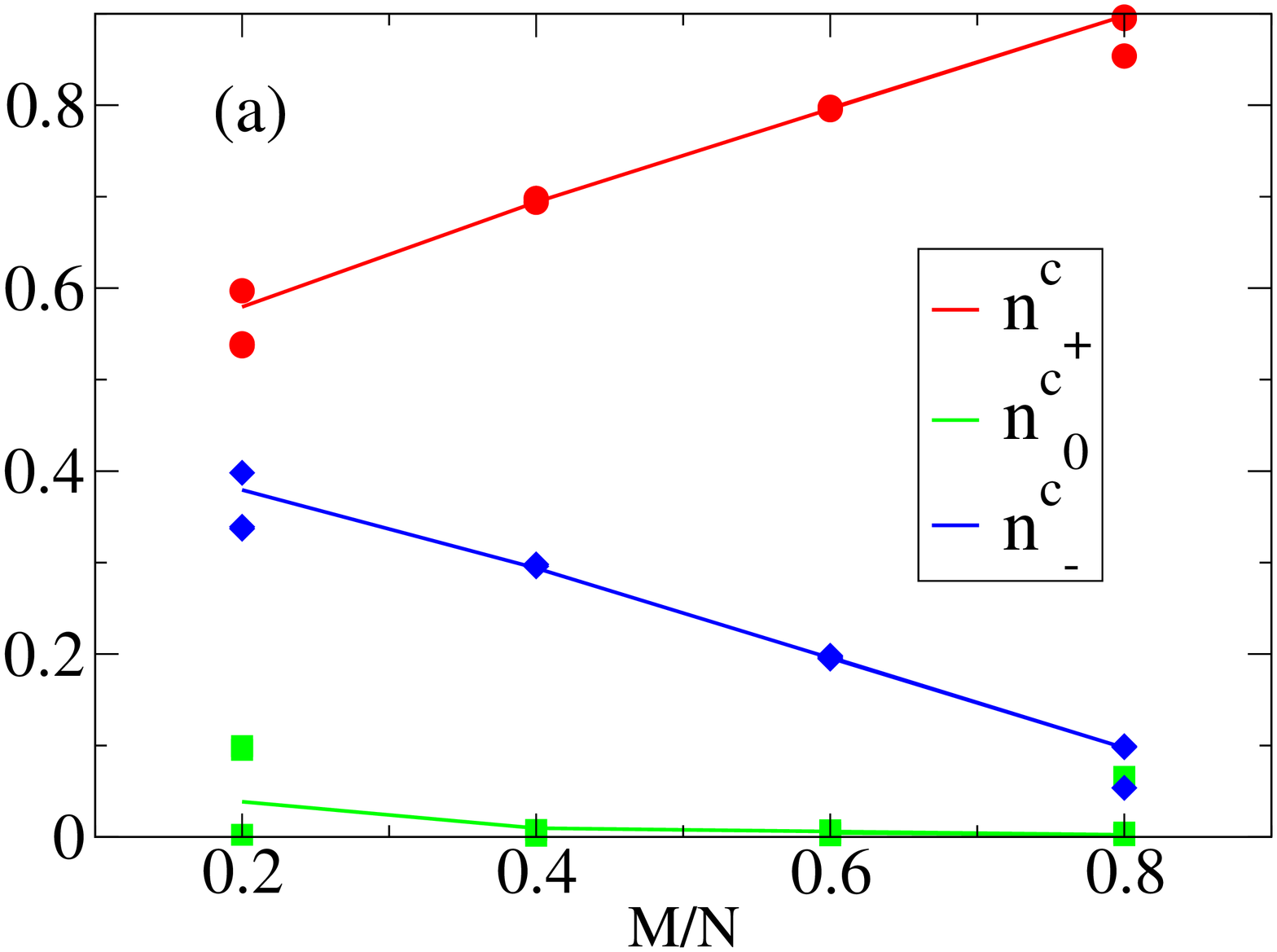}
\includegraphics[width=3.6cm, keepaspectratio, angle=-90]{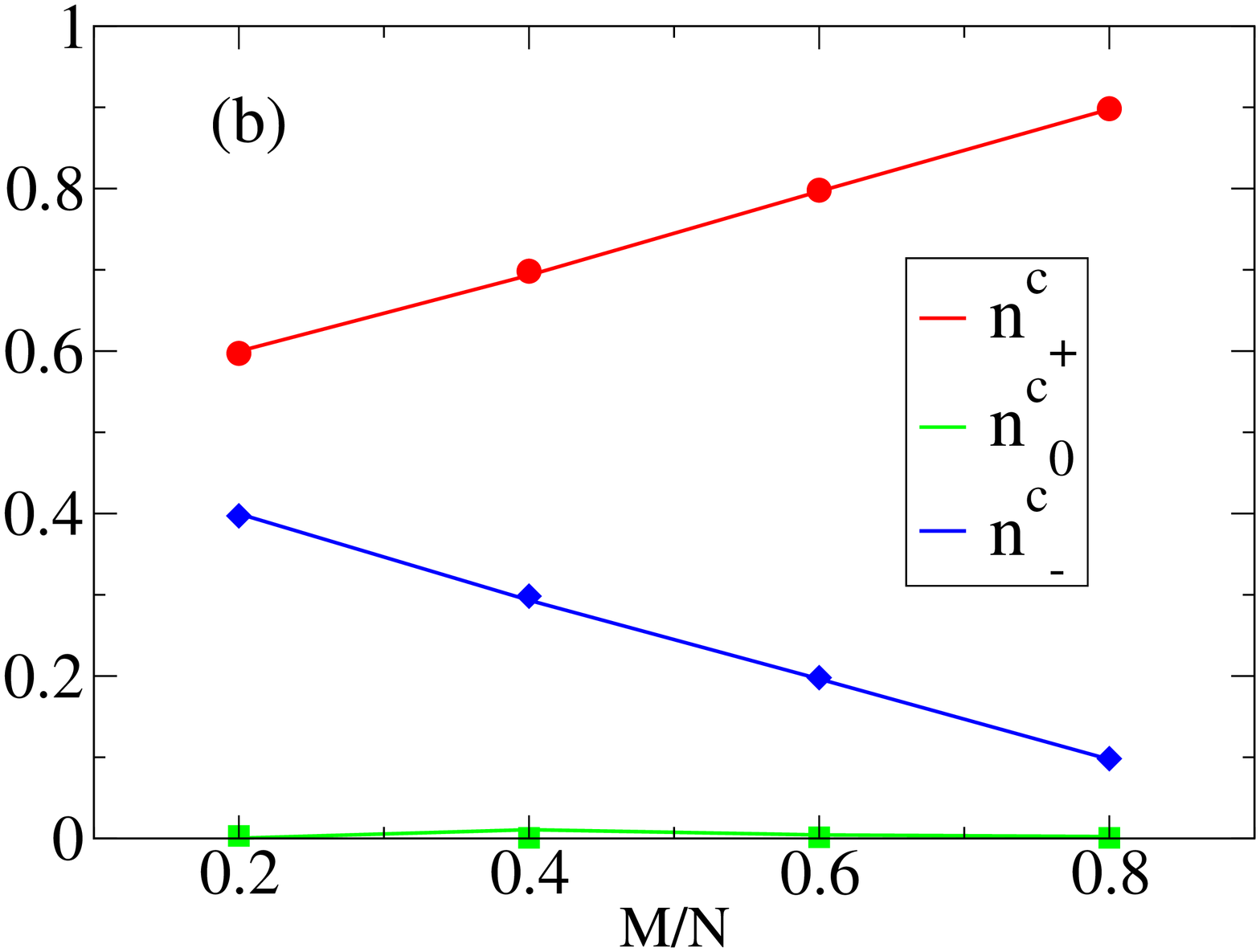}}
\caption{(Color online) A test of the Metropolis algorithm for $^{23}$Na spinor condensate in the low temperature limit.
The condensed fractions $n^c_j$, for $j=\pm, 0$, are plotted in the figure as a function of relative magnetization $M/N$ for (a) $qh^2=0$ and (b) $qh^2=0.01 \hbar^2/mL^2$. 
Particular colors denote condensate fractions in $m_F=1$ (red), $m_F=0$ (green) and $m_F=-1$ (blue) component.
Solid lines: the Bogoliubov theory, Points: Monte Carlo results. 
The total number of atoms is $N=10^5$.}
\label{bogolantiferr}
\end{figure}

In order to demonstrate the validity of the algorithm we compare Monte Carlo simulations with the exact solutions for the ideal gas in figures \ref{condensatefrac1} and \ref{condensatefrac2},
and with the approximated Bogoliubov theory for antiferromagnetic condensate in fig.\ref{bogolantiferr}. In the latter case, analytical solutions are given by the Bogoliubov transformation for antiferromagnetic interactions and are valid in the low temperature limit below the critical magnetic field~\cite{KZM_PRB}. Both comparisons are satisfactory what allows to use the algorithm in the wider range of interactions.

\subsection{Numerical results}

\begin{figure*}
\centerline{\includegraphics[width=4.8cm, keepaspectratio, angle=-90]{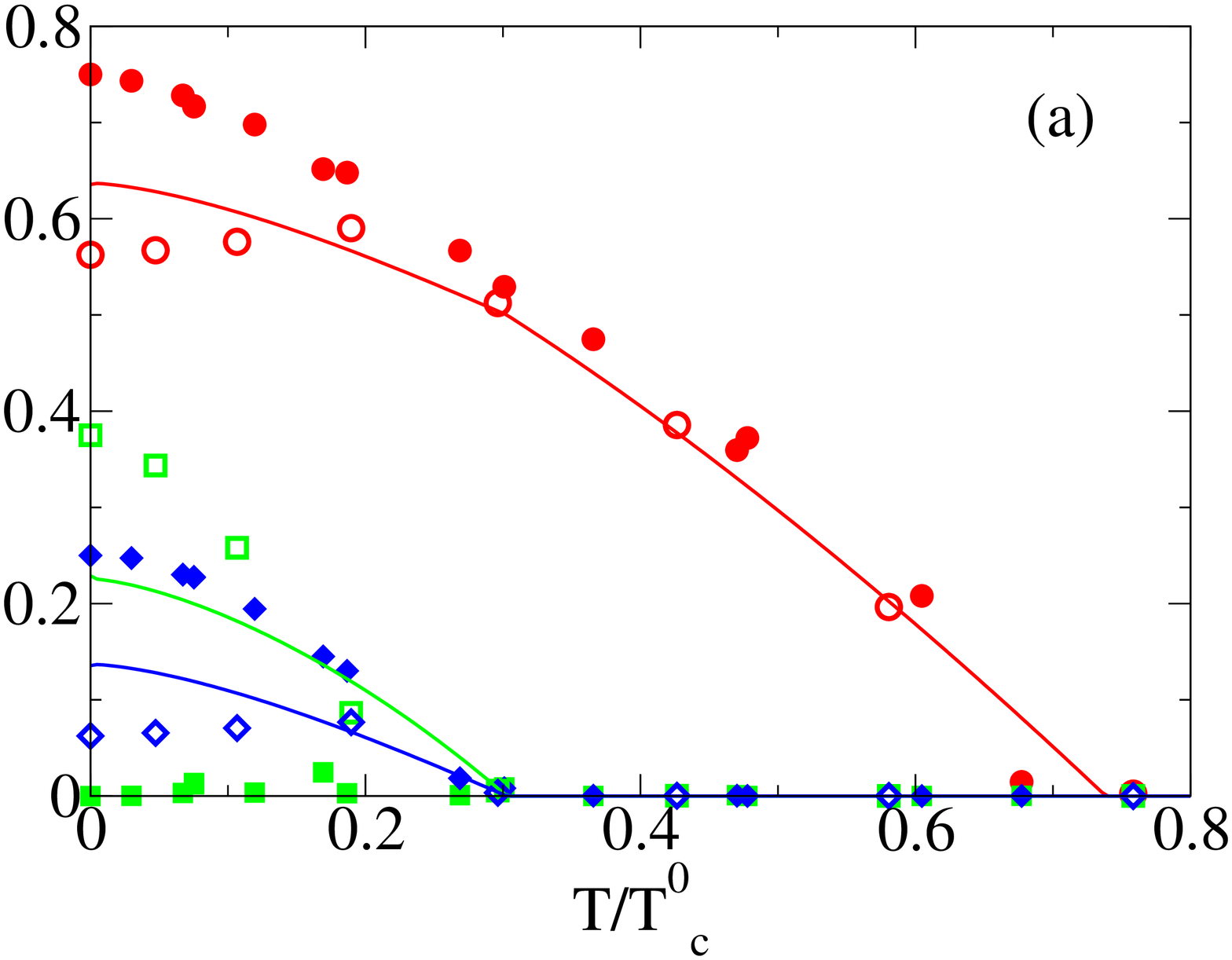}
\includegraphics[width=4.8cm, keepaspectratio, angle=-90]{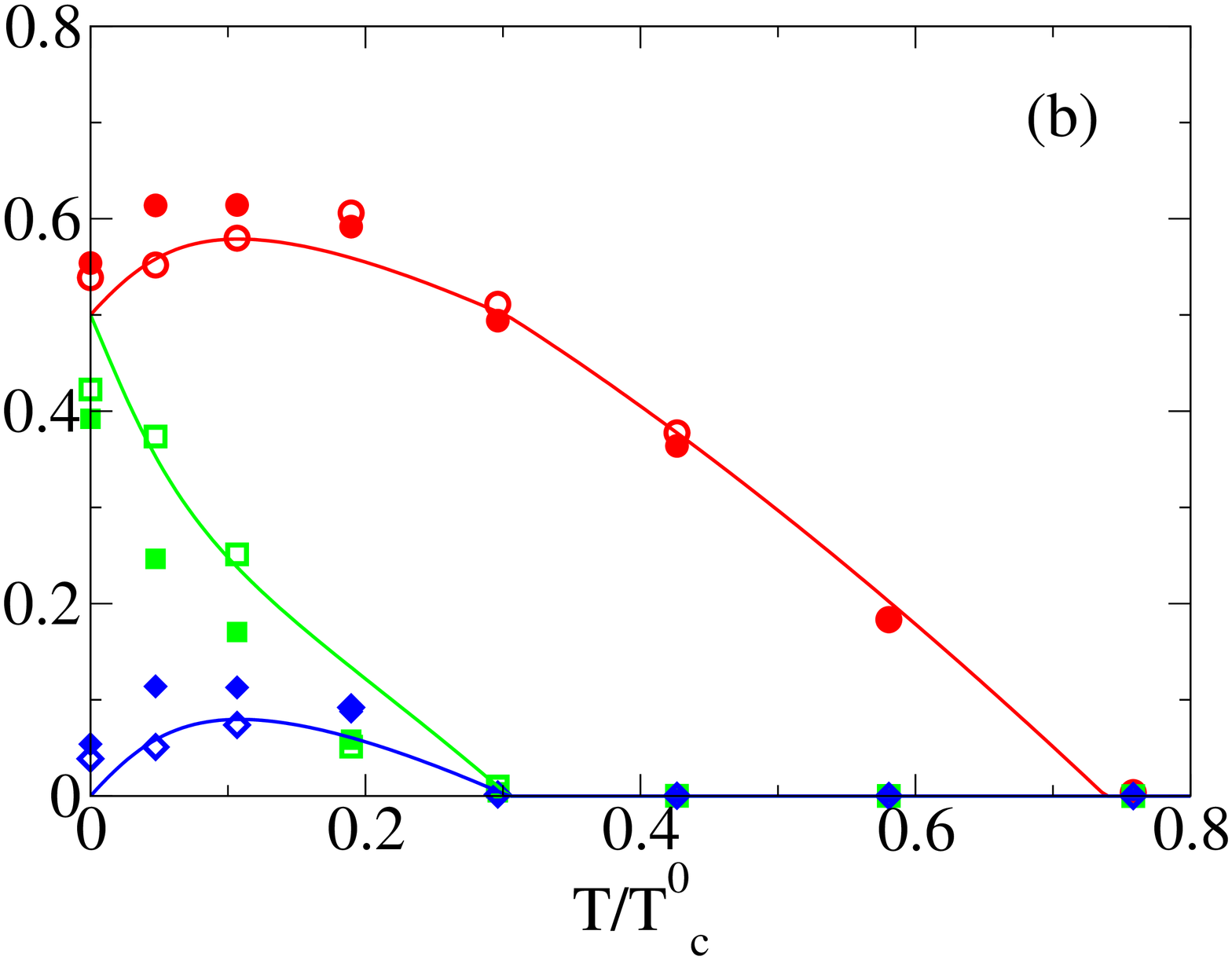}
\includegraphics[width=4.8cm, keepaspectratio, angle=-90]{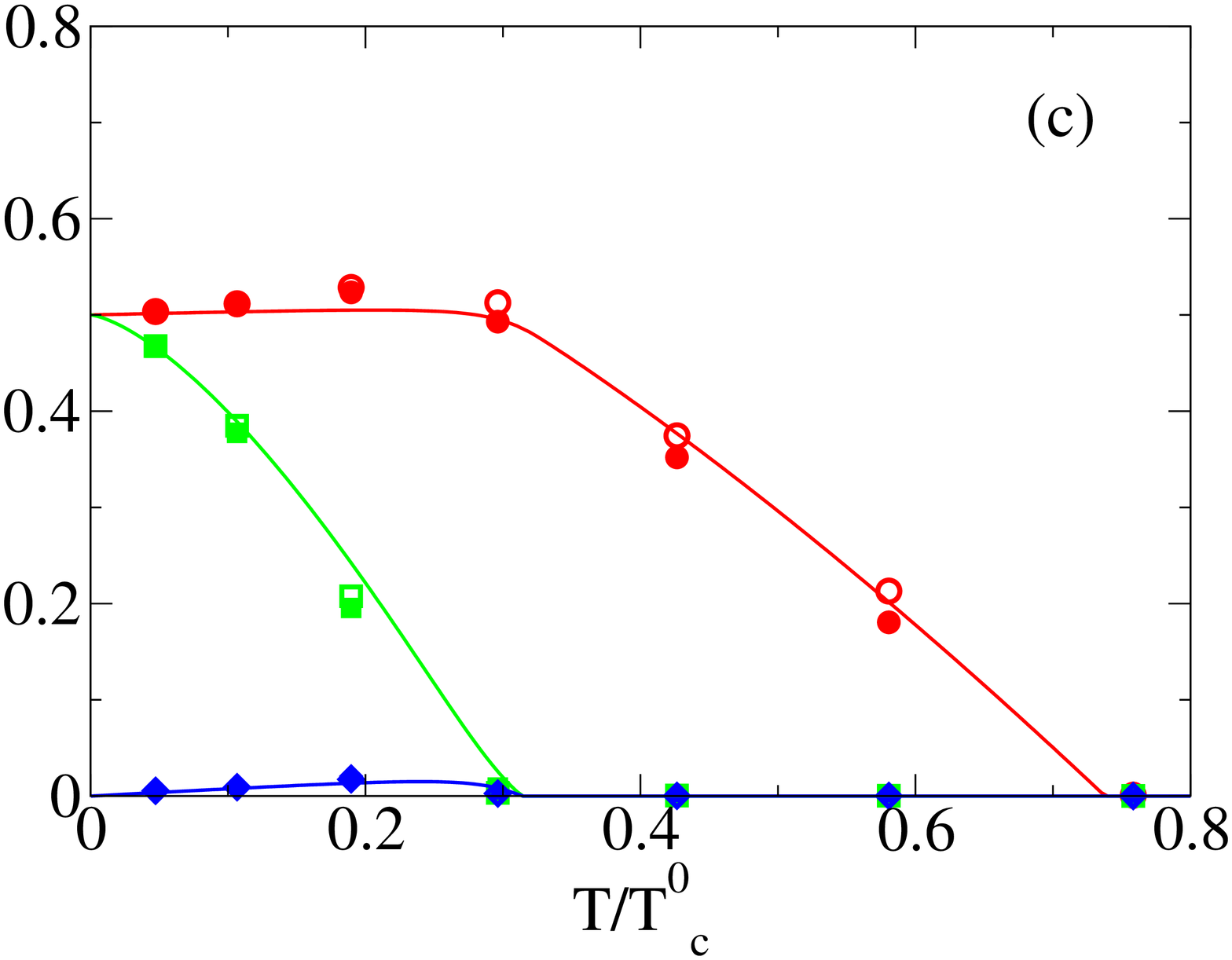}}
\caption{(Color online) Condensate fractions for $^{23}$Na atoms with antiferromagnetic interactions (filled points) and for $^{87}$Rb atoms with ferromagnetic interactions (open points).
Particular colored symbols denote condensate fractions in the $m_F=1$ component (red circles), $m_F=0$ component (green squares) and in the $m_F=-1$ component (blue diamonds).
Solid lines are results for the ideal gas.
The total number of atoms is $N=10^4$, the magnetization $M=N/2$ and values of magnetic fields are (a) $qh^2=0$, (b) $qh^2= 0.124\, \hbar^2/mL^2$ and (c) $qh^2= \hbar^2/mL^2$. }
\label{condfrac1}
\end{figure*}

\begin{figure*}
\centerline{\includegraphics[width=4.5cm, keepaspectratio, angle=-90]{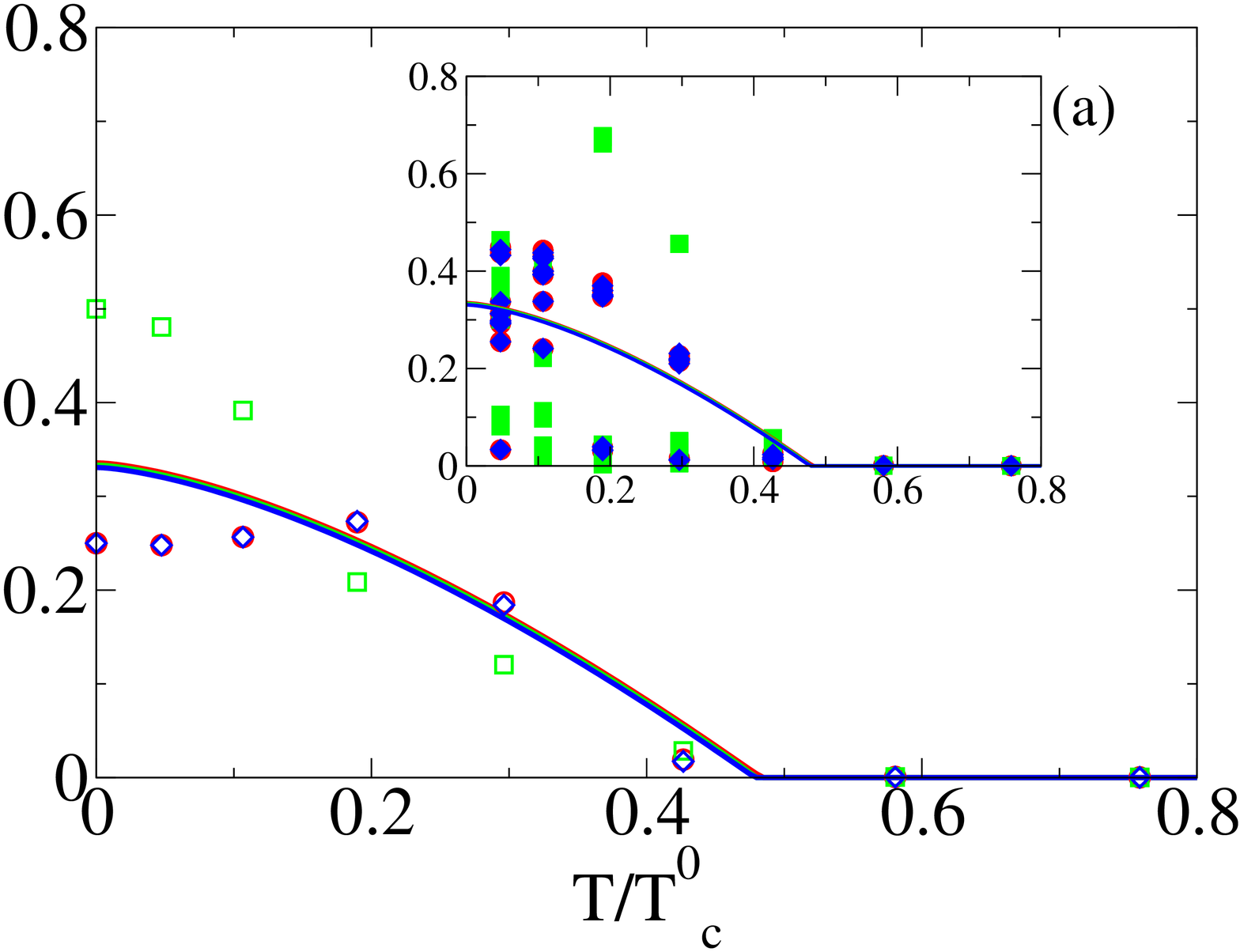}
\includegraphics[width=4.5cm, keepaspectratio, angle=-90]{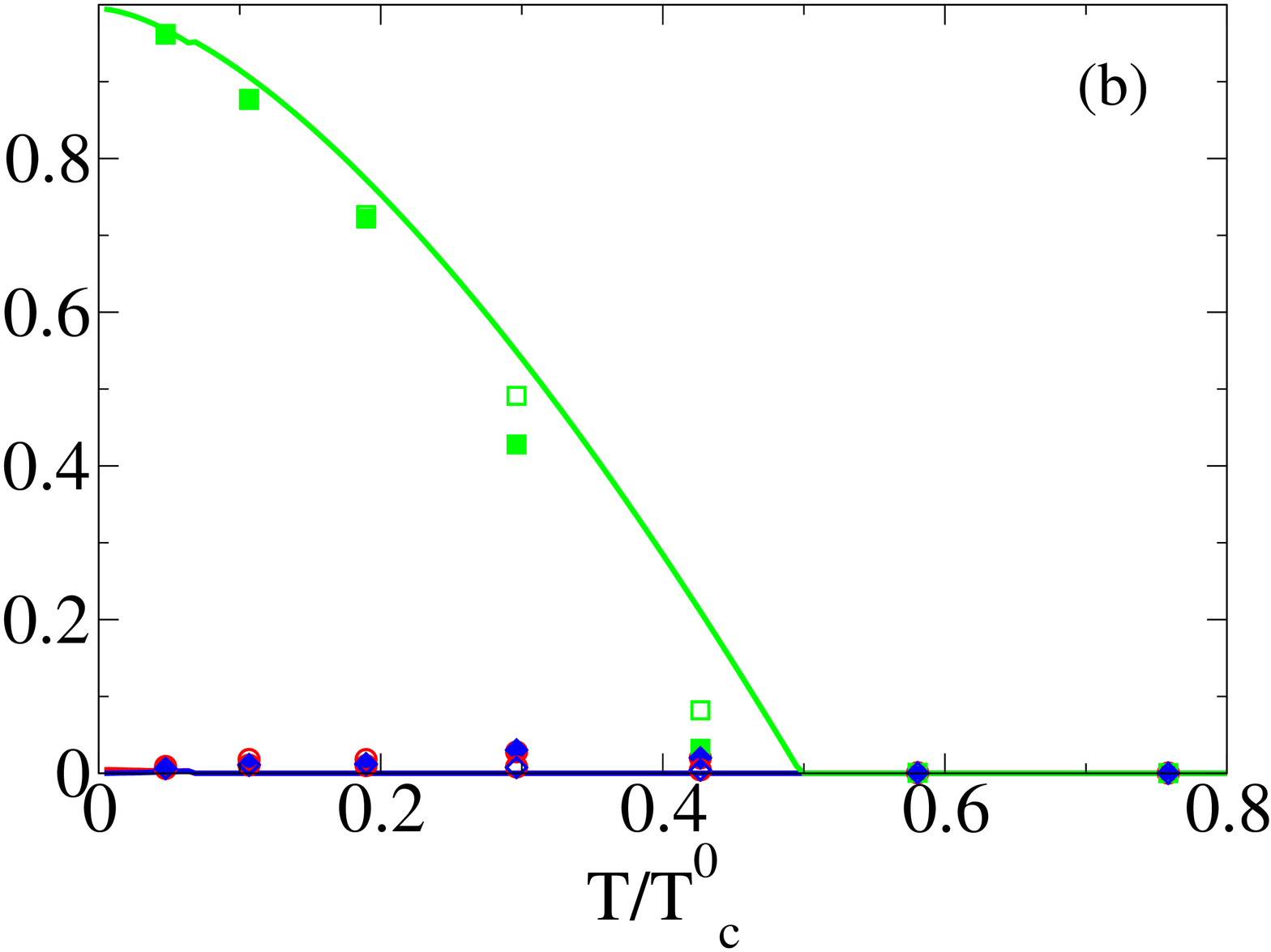}}
\caption{(Color online) The same as in fig.~\ref{condfrac1} but for magnetization $M=50$.
Condensate fractions for $^{23}$Na atoms with antiferromagnetic interactions are marked by filled points and for $^{87}$Rb atoms with ferromagnetic interactions by open points.
Particular colored symbols denote condensate fractions in the $m_F=1$ component (red circles), $m_F=0$ component (green squares) and in the $m_F=-1$ component (blue diamonds).
Solid lines are results for the ideal gas.
The total number of atoms is $N=10^4$ and the values of magnetic fields are (a) $qh^2=0$ and (b) $qh^2= \hbar^2/mL^2$.
In (a) condensed fractions for atoms with ferromagnetic interactions are presented in the main window. Noumerous points in the inset show condensate fractions for atoms with antiferromagnetic interactions that are obtained by averaging over different representations of ensemble members.
}
\label{condfrac2}
\end{figure*}

In figures \ref{condfrac1} and \ref{condfrac2} we show results of numerical simulations using the Metropolis algorithm. 
Figure \ref{condfrac1} is for the magnetization $M=N/2$, and figure \ref{condfrac2} for $M=50$. 
Condensate fractions for atoms with antiferromagnetic interactions are marked by filled points while for atoms with ferromagnetic interactions by open ones. 
Particular colored symbols denote condensate fractions in the $m_F=1$ component (red circles), $m_F=0$ component (green squares) and in the $m_F=-1$ component (blue diamonds).
Solid lines denote results for the ideal gas, added in the figures for comparison.

It clearly reveals the double phase transition that occurs in the system as it is determined by the ideal gas calculations. 
It does not seem that critical temperatures were affected very much by interactions. 
Moreover, even condensate fractions for the range of temperatures $T\in [T_{c2}, T_{c1}]$ and any magnetic field follow the ideal gas prediction. 
It is not very suprising since the system condenses in this regime like the single component gas. 	
Below the second critical temperature the condensate scenario results from the competition between spin-dependent interactions (dominant at low magnetic fields) and the quadratic Zeeman energy (dominant at large magnetic fields). The impact imposed by interactions is the most visible in the low magnetic field regime where ferromagnetic atoms condense differently than antiferromagnetic, and both do not match the ideal gas curve, see figs.~\ref{condfrac1}a and~\ref{condfrac1}b. Nevertheless, dissimilarity in populations of a given component between both interaction types is not so large.

The antiferromagnetic interaction reduces the condensate population in the $m_F=0$ component in all temperature range for magnetic fields below its critical value known from the ground state analysis~\cite{GS_theory}, see fig.~\ref{condfrac1}a. Simultaneously, the condensate fraction in the $m_F=\pm 1$ components decreases like $(T/T_c^0)^{3/2}$. 
The ferromagnetic interaction allows for condensation in all components, and populations in the lowest momentum mode may decrease or increase up to the second critical temperature depending on $m_F$. 
In the other parameters regime condensate fractions may not simply decay with the temperature but may also increase up to some temperature, reach a maximum and then decrease, see filled red points in fig.~\ref{condfrac1}b for example. This feature is also observed for the ideal gas.
In the high magnetic field regime, where the quadratic Zeeman energy dominates over the spin-dependent interaction energy, the condensate scenario matches the ideal gas prediction for both types of interactions, what can be seen in figures \ref{condfrac1}c and \ref{condfrac2}b. 

The interesting case of almost zero magnetization and zero magnetic field is presented in fig.~\ref{condfrac2}a. 
We observe strong fluctuations of condensed fractions for atoms with antiferromagnetic interactions (shown in the inset), what is not the case for ferromagnetic atoms (shown in the main window).
In the inset of fig.~\ref{condfrac2}a numerous points are obtained by averaging over different representations of ensemble members. Results of the Monte Carlo simulations strongly fluctuate, and additionally they are sensitive to the parameters of simulations (members of ensemble, or $\delta M$ for example).
Similarly as for the ideal gas, the ground state of the antiferromagnetic condensate is degenerated what gives rise to observed fluctuations. 
The phenomenon that is behind this effect is called spin fragmentation and was already investigated theoretically for the antiferromagnetic spinor condensate~\cite{spinfluctuation}.

\section{Summary}

In summary, we have studied the thermodynamics of a spin-1 Bose gas with fixed magnetization in the presence of a non-zero magnetic field.
We have given explicit expressions for the two critical temperatures and all condensate fractions for the ideal gas. 
We have shown the occurence of a new phase in the phase diagram of critical temperatures. The interacting gas was studied within the classical fields approach, that is not perturbative and includes
all nonlinear terms present in the hamiltonian. An alternative method, namely stochastic Projected Gross-Pitaevskii equation, was lately adapted to the case of a spin-1 Bose gas but for free magnetization~\cite{SGPE}. We find that interactions strongly affect the condensation scenario below the second critical temperature and for low magnetic fields. In this regime of parameters the thermodynamics of ferromagnetic and antiferromagnetic gases are different. The condensation is not affected much by interactions for values of temperatures between the two critical temperatures $T\in[T_{c1},T_{c2}]$ for all values of magnetic fields. Furthermore, the condensation is not affected by interactions in the whole temperatures range in the high magnetic field limit. Generalization to a Bose gas with arbitrary spin $F$ is straightforward. Our results open the path to study the influence of a multi-mode structure on the properties of spinor condensates, providing an interesting direction for a future work.

\acknowledgments
We thank M. Gajda and M. Matuszewski for useful discussions and O. Hul for a careful reading of the
manuscript. This work was supported by DEC-2011/03/D/ST2/01938.

\appendix

\section{Equations for condensate fractions when $T\in[0,T_{c2}]$}
\label{app:equation for condfrac}

Here we show how to obtain equations (\ref{eq:cf1-3}) for condensed fractions.

The first one is obtained by writing $N_c=N_+^c+N_0^c+N_-^c=N-N^T$ in the following
\begin{equation}
N^c=N - \left(\frac{L}{\lambda_{dB}}\right)^3 \left(2g_{\frac{3}{2}}(1)+g_{\frac{3}{2}}(e^{-2\beta qh^2})\right) \, ,
\end{equation}
then after introducing $G_{\frac{3}{2}}$ and $M^T$ it has a form
\begin{equation}
N^c=N-M^T(T) - \left(\frac{L}{\lambda_{dB}}\right)^3 G_{\frac{3}{2}}(T) \, .
\end{equation}
Knowing that $N-M=(L/\lambda_{dB})^3 G_{\frac{3}{2}}(T)$, after some algebra one finds the first equation of (\ref{eq:cf1-3}).

The second equation of (\ref{eq:cf1-3}) is just rewriting the total magnetization in terms of its condensate and thermal parts $M=(N_+^c-N_-^c)+(N_+^T-N_-^T)$, and an observation that the whole thermal part simply reduces to $M^T(T)$.

The third formula of (\ref{eq:cf1-3}) is a bit more tedious to obtain. It is a peculiar case of a more general formula, valid in any regime, that we shall prove now. Starting from the set of equations
\begin{equation}
N_+^c=\frac{z_+}{1-z_+} \, ,
\end{equation}
\begin{equation}
N_0^c=\frac{z_0}{1-z_0}=\frac{z_+z_{\eta}e^{\beta qh^2}}{1-z_+z_{\eta}e^{\beta qh^2}} \, ,
\end{equation}
\begin{equation}
N_-^c=\frac{z_-}{1-z_-}=\frac{z_+{z_{\eta}}^2}{1-z_+{z_{\eta}}^2} \, ,
\end{equation}
one rewrites
\begin{equation}
\frac{1}{N_+^c}=\frac{1}{z_+}-1 \, ,
\end{equation}
\begin{equation}
\frac{1}{N_0^c}= \frac{1}{z_+z_{\eta}e^{\beta qh^2}}-1 \, ,
\end{equation}
\begin{equation}
\frac{1}{N_-^c}= \frac{1}{z_+{z_{\eta}}^2}-1 \, ,
\end{equation}
which shows that
\begin{equation}
\frac{z_{\eta}e^{-\beta qh^2}}{N_-^c}+\frac{{z_{\eta}}^{-1}e^{-\beta qh^2}}{N_+^c}=
\frac{2}{N_0^c}+2-z_{\eta}e^{-\beta qh^2}-{z_{\eta}}^{-1}e^{-\beta qh^2} \, ,
\end{equation}
and it leads to the third equation of (\ref{eq:cf1-3}) in the limit $z_{\eta} \to e^{-\beta qh^2}$.

\section{The analytical solution of equations for condensate fractions when $T\in[0,T_{c2}]$}
\label{app:solution for condfrac}

The algebraic considerations of (\ref{eq:cf1-3}) lead to the following equation for $N_0^c$ :
\begin{equation}
{N_0^c}^3+a{N_0^c}^2+bN_0^c+c=0 \, ,
\label{eq:newsolve}
\end{equation}
where
\begin{equation}
a\equiv\frac{1-2uN_c+2\tilde{u}}{u} \, ,
\end{equation}
\begin{equation}
b\equiv -\frac{2N_c\tilde{u}-2M_{eff}u+2N_c-u{N_c}^2+uM_{eff}^2}{u} \, ,
\end{equation}
\begin{equation}
c\equiv\frac{{N_c}^2-M_{eff}^2}{u} \, ,
\end{equation}
and $u(q,h,T)\equiv \sinh(\beta qh^2)e^{-\beta qh^2}$, $\tilde{u}(q,h,T)\equiv \cosh(\beta qh^2)e^{-\beta qh^2}$.

The quadratic Zeeman effect transformed the equation for $N_0^c$ which was a second degree polynomial for the zero magnetic field, into a third degree polynomial. That polynomial has three roots. One needs to select, among those solutions, the only one that is physical: real, non-negative, and with values between $0$ and $N$. To avoid numerical difficulties, one can find analytical solutions of this equation using for instance Cardan's method, and select the one that has the proper limit when $qh^2 \to 0$.
To do it one defines:
\begin{equation}
X\equiv N_0^c+\frac{a}{3} \, ,
\end{equation}
which allows to put the polynomial into the form:
\begin{equation}
X^3+\tilde{p}X+\tilde{q}=0
\end{equation}
with
\begin{equation}
\tilde{p}\equiv b-\frac{a^2}{3} \, ,
\end{equation}
\begin{equation}
\tilde{q}\equiv\frac{a}{27}(2a^2-9b)+c.
\end{equation}
Then, one writes:
\begin{equation}
X\equiv u+v \, ,
\end{equation}
and notices that $u^3$ and $v^3$ are solutions of 
\begin{equation}
X^2+\tilde{q}X-\tilde{p}^3/27=0 \, .
\end{equation}
Then one introduces 
\begin{equation}
\Delta \equiv \frac{27\tilde{q}^2+4\tilde{p}^3}{27} \, .
\end{equation}
Numerically, it appears that $\Delta<0$ and $\tilde{p}<0$, which means that there are three solutions:
\begin{equation}
X_k=2\sqrt{-\tilde{p}/3}\cos\left(\frac{1}{3}\arccos\left(-\frac{\tilde{q}}{2}\sqrt{\frac{27}{-\tilde{p}^3}}\right)+\frac{2k\pi}{3}\right) \, ,
\end{equation}
where $k \in \{0,1,2\}$.
To find ${N_0^c}_k$, one has to keep in mind that
\begin{equation}
{N_0^c}_k=X_k-\frac{a}{3}.
\end{equation}
Eventually, one finds that ${N_0^c}_1$ should always be selected for $N_0^c$ because it is the only solution that gives the appropriate limit when $qh^2 \to 0$. Then, having $N_0^c$, one can easily calculate $N_+^c$, $N_-^c$ with the first and second equations of (\ref{eq:cf1-3}).

\section{How to obtain the phase diagram}
\label{diagrammephaseh}

In this appendix we explain how to compute numerically the transition temperatures $T_{c1}$ and $T_{c2}$ for any fixed magnetization.

There is an additional difficulty to compute $T_{c1}$ compared to the case when $h=0$, since its definition involves $z_{\eta c1}\equiv
 z_{\eta}(T_{c1})$, which is unknown. One can determine $z_{\eta c1}$ from the constant of motion $M/N$, but to do so, one has to know the value of $T_{c1}$, 
as can be seen from the following sets of equations :
\begin{equation}
T_{c1}\equiv C\left(\frac{N}{F_{\frac{3}{2}}^+(T_{c1},z_{\eta c1})}\right)^{\frac{2}{3}},
\label{eq:T11}
\end{equation}
\begin{equation}
\frac{M}{N}=\frac{g_{\frac{3}{2}}(1)-g_{\frac{3}{2}}(z_{\eta c1}^2)}{g_{\frac{3}{2}}(1)+g_{\frac{3}{2}}(e^{\frac{qh^2}{k_BT_{c1}}}z_{\eta c1})+g_{\frac{3}{2}}(z_{\eta c1}^2)},
\label{eq:T12}
\end{equation}
if $qh^2 \le \eta$, and
\begin{equation}
T_{c1}\equiv C\left(\frac{N}{F_{\frac{3}{2}}^0(T_{c1},z_{\eta c1})}\right)^{\frac{2}{3}},
\label{eq:T11bis}
\end{equation}
\begin{equation}
\frac{M}{N}=\frac{g_{\frac{3}{2}}(e^{-\frac{qh^2}{k_BT_{c1}}}{z_{\eta c1}}^{-1})-g_{\frac{3}{2}}(e^{-\frac{qh^2}{k_BT_{c1}}}z_{\eta c1})}{g_{\frac{3}{2}}(e^{-\frac{qh^2}{k_BT_{c1}}}{z_{\eta c1}}^{-1})+g_{\frac{3}{2}}(1)+g_{\frac{3}{2}}(e^{-\frac{qh^2}{k_BT_{c1}}}z_{\eta c1})},
\label{eq:T12bis}
\end{equation}
if $qh^2 \ge \eta$. There are no further independent equations available for those two quantities, so they have to be solved in a self-consistent way.

Let us suppose that the magnetization is such that the system is in the area where $qh^2 \le \eta$. One knows the value of $T_{c1}$ without magnetic field, and one can sensibly expect that if a magnetic field is switched on, the critical temperature will be of the same order of magnitude as it used to be, so one puts the value of $T_{c1}$ in the absence of any field $qh^2=0$ to compute the value of $z_{\eta c1}$, and then puts this value into (\ref{eq:T11}) to compute the corrected value of $T_{c1}$, that can be used in (\ref{eq:T12}) to compute $z_{\eta c1}$.
Those operations should be performed as many times as needed to make the effect of the wrong initial value disappear. The convergence is fast, and after few steps one is close to the fixed point for $T_{c1}$.

One can proceed in the same way using (\ref{eq:T11bis}) and (\ref{eq:T12bis}) if $qh^2 \ge \eta$, but how is it possible to know at once if one is in this case or in the other? One does not know it, but it is of no importance whatsoever if one uses a little trick familiar to chemists. At the beginning one makes some assumption, and then checks if the computed value $z_{\eta c1}$ is consistent with this guess. If not, then it means that the assumption was wrong, and that the system is in the other area. One should begin calculations again. Numerical problems can occur if one is really near to the border between the two areas, so one needs to be careful.

Eventually, to find the second critical temperature, one computes the only solution of the equation in $T_{c2}$ 
\begin{equation}
T_{c2}- C\left(\frac{N-M}{M^{T}(qh^2,T_{c2})+3g_{\frac{3}{2}}(e^{\frac{-2qh^2}{k_BT_{c2}}})}\right)^{\frac{2}{3}}=0
\end{equation}
if $qh^2 \le \eta$, or
\begin{equation}
T_{c2}-C\left(\frac{M}{g_{\frac{3}{2}}(1)-g_{\frac{3}{2}}(e^{\frac{-2qh^2}{k_BT_{c2}}})}\right)^{\frac{2}{3}}=0
\end{equation}
if $qh^2 \ge \eta$. The bisection method allows to find this value with the requested accuracy, taking $0$ for the lower bound and the temperature $T_{c1}$ for the upper bound.



\begin{thebibliography}{99}

\bibitem{quantum_magnetism} 
D. M. Stamper-Kurn and M. Ueda, {Rev. Mod. Phys.} {\bf 85}, 1191 (2013).

\bibitem{superfluidity} 
Y. Kawaguchi and M. Ueda, {Phys. Rep.} {\bf 520}, 253 (2012).

\bibitem{strong_correlations} 
Z. Zhang and L.-M. Duan, {Phys. Rev. Lett.} {\bf 111}, 180401 (2013).

\bibitem{spin_mixing} 
M.-S. Chang, Q. Qin, W. Zhang, L. You and M. S. Chapman, {Nature Physics} {\bf 1}, 111 (2005).
W. Zhang, D. L. Zhou, M.-S. Chang, M. S. Chapman and L. You, {Phys. Rev. A} {\bf 72}, 0.13602 (2005).
C. S. Gerving, T. M. Hoang, B. J. Land, M. Anquez, C. D. Hamley and M. S. Chapman, {Nature Communications} {\bf 3}, 1169 (2012).

\bibitem{spin_squeezing} 
T. M. Hoang, C. S. Gerving, B. J. Land, M. Anquez, C. D. Hamley and M. S. Chapman, {Phys. Rev. Lett.} {\bf 111}, 090403 (2013).

\bibitem{massive_entanglement} 
C. D. Hamley, C. S. Gerving, T. M. Hoang, E. M. Bookjans and M. S. Chapman, {Nature Physics} {\bf 8}, 305 (2012).

\bibitem{StamperKurn}
J. Guzman, G.-B. Jo, A. N. Wenz, K. W. Murch, C. K. Thomas, and D. M. Stamper-Kurn, Phys. Rev. A {\bf 84}, 063625 (2011).

\bibitem{GS_experiment}
D. Jacob, L. Shao, V. Corre, T. Zibold, L. De Sarlo, E. Mimoun, J. Dalibard and F. Gerbier, {Phys. Rev. A} {\bf 86}, 061601 (2012).
J. Jiang, L. Zhao, M. Webb, and Y. Liu, Phys. Rev. A {\bf 90}, 023610 (2014).

\bibitem{Isochima}
T. Isoshima, T. Ohmi and K. Machida, {J. Phys. Soc. Jpn.} {\bf 69}, 3864 (2000).

\bibitem{Kao}
Y. M. Kao and T. F. Jiang, {Eur. Phys. J. D} {\bf 40}, 263 (2006).

\bibitem{Zhang}
W. Zhang, S. Yi and L. You, {Phys. Rev. A} {\bf 70}, 043611 (2004). 

\bibitem{Laburthe}
B. Pasquiou, G. Bismut, Q. Beaufils, A. Crubellier, E. Marechal, P. Pedri, L. Vernac, O. Gorceix and B. Laburthe-Tolra, Phys. Rev. A {\bf 81}, 042716 (2010).
B. Pasquiou, E. Marechal, L. Vernac, O. Gorceix and B. Laburthe-Tolra, Phys. Rev. Lett. {\bf 108}, 045307 (2012).





\bibitem{CFM}
Y. Kagan, B. V. Svistunov, {Phys. Rev. Lett.} {\bf 79}, 3331 (1997).
M. J. Davis, S. A. Morgan and K. Burnett {Phys. Rev. Lett.} {\bf 87}, 160402 (2001).
A. Sinatra, C. Lobo, Y. Castin, {Phys. Rev. Lett.} {\bf 87}, 210404 (2001).
K. G\'oral, M. Gajda, K. Rz\k{a}\.zewski, {Opt. Express} {\bf 8}, 92 (2001).
M. Brewczyk, P. Borowski, M. Gajda, K. Rz\k{a}\.zewski, {J. Phys. B} {\bf 37} 2725 (2004).

\bibitem{metropolis}
N. Metropolis et al., {J. Chem. Phys.} {\bf 21} 1087 (1953).

\bibitem{Optc}
E. Witkowska, M. Gajda, K. Rz\k{a}\.zewski, {Optics Communications} {\bf 283} 671-675 (2010).

\bibitem{vortex}
C. Lobo, A. Sinatra and Y. Castin, {Phys. Rev. Lett.} {\bf 92}, 020403 (2004).
T. Karpiuk, M. Brewczyk, M. Gajda, K. Rz\k{a}\.zewski, {J. Phys. B: At. Mol. Opt. Phys.} {\bf 42} No 9 095301 (2009).
C. N. Weiler, T. W. Neely, D. R. Scherer, A. S. Bradley, M. J. Davis, and B. P. Anderson, {Nature} {\bf 455}, 948 (2008).

\bibitem{Tcshift}
M. J. Davis, S. A. Morgan, {Phys. Rev. A} {\bf 68}, 053615 (2003).

\bibitem{sqthermal}
A. Sinatra, E. Witkowska, J.-C. Dornstetter, Yun Li, and Y. Castin, {Phys. Rev. Lett.} {\bf 107}, 060404 (2011).
A. Sinatra, Y. Castin, E. Witkowska, {Europhysics Letters} {\bf 102} 40001 (2013).

\bibitem{solitons}
E. Witkowska, P. Deuar, M. Gajda, K. Rz\k{a}\.zewski, {Phys. Rev. Lett.} {\bf 106}, 135301 (2011).
T. Karpiuk, P. Deuar, P. Bienias, E. Witkowska, K. Pawlowski, M. Gajda, K. Rz\k{a}\.zewski, M. Brewczyk, {Phys. Rev. Lett.} {\bf 109}, 205302 (2012).

\bibitem{ref_TH}
{Finite Temperature and Non-Equilibrium Dynamics}, S. Gardiner, N. Proukakis, M. Davis, M. Szymanska editors, Cold Atom Series, 1, Imperial College Press, London  2013.

\bibitem{GS_theory}
J. Stenger, S. Inouye, D. M. Stamper-Kurn, H.-J. Miesner, A. P. Chikkatur and W. Ketterle, {Nature} {\bf 396}, 345 (1998).
W. X. Zhang, S. Yi and L. You, {New J. Phys.} {\bf 5}, 77 (2003).
K. Murata, H. Saito and M. Ueda, {Phys. Rev. A} {\bf 75}, 013607 (2007).
M. Matuszewski, T. J. Alexander and Y. S. Kivshar, {Phys. Rev. A} {\bf 80}, 023602 (2009).

\bibitem{Matuszewski_AF} 
M. Matuszewski, T. J. Alexander, and Y. S. Kivshar, {Phys. Rev. A} {\bf 78}, 023632 (2008).

\bibitem{SGPE}
A. S. Bradley and P. B. Blakie,  Phys. Rev. A {\bf 90}, 023631 (2014).

\bibitem{other_meth}
N. T. Phuc, Y. Kawaguchi and M. Ueda, {Phys. Rev. A} {\bf 84}, 043645 (2011).
Y. Kawaguchi, N. T. Phuc and P. B. Blakie, {Phys. Rev. A} {\bf 85}, 053611 (2012).

\bibitem{our_EPJST}
E. Witkowska, M. Gajda, K. Rz\k{a}\.zewski, {Phys. Rev. A} {\bf 79} 033631 (2009).
A.Sinatra, E. Witkowska, Y. Castin, {Eur. Phys. J. Special Topics} {\bf 203}, 87-116 (2012).

\bibitem{KZM_PRB}
E. Witkowska, J. Dziarmaga, T. \'Swis\l{}ocki, M. Matuszewski, {Phys. Rev. B} {\bf 88}, 054508 (2013).

\bibitem{spinfluctuation}
L. DeSarlo, L. Shao, V. Corre, T. Zibold, D. Jacob, J. Dalibard, F. Gerbier, {New J. Phys.} {\bf 15} 113039 (2013).


\end{thebibliography}
\end{document}